\documentclass[10pt,draftcls,onecolumn]{IEEEtran}
\usepackage[latin9]{inputenc}
\usepackage{color}
\usepackage{amsmath}
\usepackage{amsthm}
\usepackage{amssymb}
\usepackage{graphicx}
\usepackage[unicode=true,pdfusetitle,
 bookmarks=true,bookmarksnumbered=true,bookmarksopen=false,
 breaklinks=false,pdfborder={0 0 0},pdfborderstyle={},backref=false,colorlinks=true]
 {hyperref}
\hypersetup{
 pdfborderstyle={},pdfborderstyle={},pdfpagelayout=OneColumn,pdfnewwindow=true,pdfstartview=XYZ,plainpages=false,linkcolor=blue,urlcolor=blue,citecolor=red,anchorcolor=blue,linkcolor=blue,urlcolor=blue,citecolor=red,anchorcolor=blue}

\makeatletter
\theoremstyle{plain}
\newtheorem{thm}{\protect\theoremname}
\theoremstyle{definition}
\newtheorem{defn}[]{\protect\definitionname}
\theoremstyle{remark}
\newtheorem{rem}[]{\protect\remarkname}
\theoremstyle{plain}
\newtheorem{lem}[]{\protect\lemmaname}

\usepackage{latexsym}
\usepackage{bm}
\usepackage{wrapfig}
\usepackage{fancybox}

\usepackage{bm}
\usepackage{epstopdf}
\usepackage{float}
\usepackage{subfloat}
\usepackage{array}
\usepackage{tabularx}
\usepackage{multirow}
\usepackage{tikz}
\usepackage{algorithmic}

\tikzstyle{arw}=[->,>=latex]
\tikzstyle{node}=[draw,rectangle,rounded corners, minimum width=1cm,minimum height =.75 cm]

\usepackage{color}

\usepackage{adjustbox}

\usepackage{algorithmic}

\providecommand{\definitionname}{Definition}
\providecommand{\lemmaname}{Lemma}

\providecommand{\theoremname}{Theorem}

\providecommand{\definitionname}{Definition}
\providecommand{\lemmaname}{Lemma}
\providecommand{\remarkname}{Remark}
\providecommand{\theoremname}{Theorem}

\providecommand{\definitionname}{Definition}
\providecommand{\lemmaname}{Lemma}
\providecommand{\remarkname}{Remark}
\providecommand{\theoremname}{Theorem}

\newtheorem{property}{Property}

\providecommand{\definitionname}{Definition}
\providecommand{\lemmaname}{Lemma}
\providecommand{\remarkname}{Remark}
\providecommand{\theoremname}{Theorem}
\usepackage{amsfonts}
\usepackage{amsopn}
\usepackage{bbm}

\def\tR{\tilde{R}}

\def\be{\begin{equation}}
\def\ee{\end{equation}}
\def\bes{\begin{equation*}}
\def\ees{\end{equation*}}
\def\beq{\begin{eqnarray}}
\def\eeq{\end{eqnarray}}
\def\beqs{\begin{eqnarray*}}
\def\eeqs{\end{eqnarray*}}

\def\mx{{\mathcal X}}

\def\e{\mathbb{E}}

\def\tv#1{\left\|#1\right\|_{TV}}
\def\apx#1{\stackrel{#1}{\approx}}

 \def\clap#1{\hbox to 0pt{\hss#1\hss}}

\allowdisplaybreaks

\providecommand{\definitionname}{Definition}
\providecommand{\lemmaname}{Lemma}
\providecommand{\remarkname}{Remark}
\providecommand{\theoremname}{Theorem}

\makeatother

\providecommand{\definitionname}{Definition}
\providecommand{\lemmaname}{Lemma}
\providecommand{\remarkname}{Remark}
\providecommand{\theoremname}{Theorem}

\begin{document}

\title{Generalized Common Informations: \\
Measuring Commonness by the Conditional Maximal Correlation}

\author{Lei Yu, Houqiang Li, \textit{Senior Member, IEEE}, and Chang Wen
Chen, \textit{Fellow, IEEE}\thanks{Lei Yu is with the Department of Electrical and Computer Engineering,
National University of Singapore, Singapore (e-mail: leiyu@nus.edu.sg).
This work was done when he was at University of Science and Technology
of China. Houqiang Li is with the Department of Electronic Engineering
and Information Science, University of Science and Technology of China,
Hefei, China (e-mail: lihq@ustc.edu.cn). Chang Wen Chen is with Department
of Computer Science and Engineering, State University of New York
at Buffalo, Buffalo, NY, USA (e-mail: chencw@buffalo.edu).}}
\maketitle
\begin{abstract}
In literature, different  common informations were defined by Gács
and Körner, by Wyner, and by Kumar, Li, and Gamal, respectively. In
this paper, we  define two generalized versions of common informations,
named \emph{approximate }and\emph{ exact information-correlation functions},
by exploiting the conditional maximal correlation as a commonness
or privacy measure. These two generalized common informations encompass
the notions of Gács-Körner's, Wyner's, and Kumar-Li-Gamal's common
informations as special cases. Furthermore, to give operational characterizations
of these two generalized common informations, we also study the problems
of private sources synthesis and common information extraction, and
show that the information-correlation functions are equal to  the
minimum rates of commonness needed to ensure that some conditional
maximal correlation constraints are satisfied for the centralized
setting versions of these problems. As a byproduct,  the conditional
maximal correlation has been studied as well.

\end{abstract}


\section{Introduction}

Common information, as an information measure on the common part between
two random variables, was first investigated by Gács and Körner \cite{G=00003D00003D0000E1cs}
 in content of distributed common information extraction problem:
extracting a same random variable from each of two sources individually.
The common information of the sources is defined by the maximum information
of the random variable that can be extracted from them. For correlated
memoryless sources $X,Y$ (taken from finite alphabets), \cite{G=00003D00003D0000E1cs}
shows that the Gács-Körner common information between them is
\begin{equation}
C_{GK}(X;Y)={\displaystyle \sup_{f,g:f\left(X\right)=g\left(Y\right)}H(f\left(X\right))}.\label{eq:-77}
\end{equation}
It also can be expressed as
\begin{equation}
C_{GK}(X;Y)={\displaystyle \inf_{P_{U|XY}:C_{GK}(X;Y|U)=0}I(XY;U)},\label{eq:-76}
\end{equation}
(the proof of \eqref{eq:-76} is given in Appendix \ref{sec:Proof-of-Equation}),
where
\begin{equation}
C_{GK}(X;Y|U):={\displaystyle \sup_{f,g:f\left(X,U\right)=g\left(Y,U\right)}H(f\left(X,U\right)|U)}\label{eq:-78}
\end{equation}
denotes the conditional common information between $X,Y$ given $U$.
 The constraint $C_{GK}(X;Y|U)=0$ in \eqref{eq:-76} implies all
the common information between $X,Y$ is contained in $U$.

Wyner \cite{Wyner} studied distributed source synthesis (or distributed
source simulation) problem, and defined common information in a different
way. Specifically, he defined common information as the minimum information
rate needed to generate sources in a distributed manner with asymptotically
vanishing normalized relative entropy between the induced distribution
 and some target joint distribution. Given a target distribution
$P_{XY}$, this  common information is proven to be
\begin{equation}
C_{W}(X;Y)={\displaystyle \inf_{P_{U|XY}:X\rightarrow U\rightarrow Y}I(XY;U)}.\label{eq:-24-2}
\end{equation}
Furthermore, as a related problem, the problem of \emph{exactly} generating
target sources was studied by Kumar, Li, and Gamal recently \cite{Kumar}.
The notion of exact common information (rate) (denoted as $K_{KLG}(X;Y)$)
is introduced, which is defined to be the minimum code rate to ensure
the induced distribution is exactly (instead approximately) same to
some target joint distribution. By comparing these common informations,
it is easy to show that $C_{GK}(X;Y)\leq I(X;Y)\leq C_{W}(X;Y)\leq K_{KLG}(X;Y)\leq H(XY)$.

Observe that in the definitions of Gács-Körner and Wyner common informations,
different dependency constraints are used. Gács-Körner common information
requires the common variable $U$ to be some function of each of the
sources (or equivalently, there is no conditional common information
given $U$); while Wyner common information requires the sources conditionally
independent given the common variable $U$. These two constraints
are closely related to an important dependency measure, \emph{Hirscbfeld-Gebelein-Renyi
maximal correlation} (or simply \emph{maximal correlation}). This
correlation measures the maximum (Pearson) correlation between square
integrable real-valued random variables generated by the individual
random variables.  According to the definition, maximal correlation
is invariant on bijective mappings (or robust to bijective transform),
hence it reveals some kind of intrinsic dependency between two sources.
This measure was first introduced by Hirschfeld \cite{Hirschfeld}
and Gebelein \cite{Gebelein}, then studied by Rényi \cite{R=00003D00003D0000E9nyi},
and recently it has been exploited to some interesting problems of
information theory, such as measure of non-local correlations \cite{Beigi},
maximal correlation secrecy \cite{Li}, converse result of distributed
communication \cite{Yu}, etc. Furthermore, maximal correlation also
indicates the existence of Gács-Körner or Wyner common information:
There exists Gács-Körner common information between two sources if
and only if the maximal correlation between them equals one; and there
exists Wyner common information between two sources if and only if
the maximal correlation between them is positive.

The common informations proposed by Gács and Körner and by Wyner (or
by Kumar, Li, and Gamal) are defined in two different problems: distributed
common information extraction and distributed source synthesis. In
these problems, the common informations are defined from different
points of view. One attempt to unify them can be found in \cite{Kamath},
where Kamath and Anantharam converted common information extraction
problem into a special case of distributed source synthesis problem
by specifying the synthesized distribution to be that of the common
randomness. In this paper, we attempt to give another unification
of the existing common informations. Specifically, we unify and generalize
the Gács-Körner and Wyner common informations by defining a generalized
common information, \emph{(approximate) information-correlation function}.
In this generalized definition, the conditional maximal correlation
(the conditional dependency of the sources given the common randomness)
is exploited to measure the privacy (or commonness), and the mutual
information is used to measure the information amount of such common
randomness. The Gács-Körner common information and Wyner common information
are two special and extreme cases of our generalized definition with
correlation respectively being 0  and $1^{-}$\footnote{$1^{-}$ implies the correlation approaching 1 from the left.},
and hence both of them can be seen as hard-measures of common information.
However, in our definition, correlation could be any number between
0 and 1, hence our definition gives a soft-measure of common information.
Our results give a more comprehensive answer to the classic problem:
What is the common information between two correlated sources? Furthermore,
similarly we also unify and generalize the Gács-Körner and Kumar-Li-Gamal
common informations into another generalized common information, \emph{(exact)
information-correlation function}. To give an operational interpretation
of the approximate and exact generalized common informations, we also
study common information extraction problem and private sources synthesis
problem, and show that the information-correlation functions correspond
to the minimum achievable rates under privacy constraints for the
centralized case of each problem.

The rest of this paper is organized as follows. Section II summarizes
definitions and properties of maximal correlation. Section III defines
information-correlation function and provides the basic properties.
Sections IV and V investigate the private sources synthesis problem
and common information extraction problem respectively. Finally, Section
VI gives the concluding remarks.

\subsection{\textit{\emph{Notation and Preliminaries}}}

We use $P_{X}(x)$ to denote the probability distribution of random
variable $X$, which is also shortly denoted as $P_{X}$ or $P(x)$.
We also use $P_{X}$ and $Q_{X}$ to denote different probability
distribution with common alphabet $\mathcal{X}.$ We use $P_{X}^{U}$
to denote the uniform distribution over the set $\mathcal{X}$, unless
otherwise stated. We use $f_{P}$ or $f_{Q}$ to denote a quantity
or operation $f$ that is defined on pmf $P$ or $Q$. The total variation
distance between two probability measures $P$ and $Q$ with common
alphabet is defined by
\begin{equation}
{\displaystyle \Vert P-Q\Vert_{TV}:=\sup_{A\in\mathcal{F}}|P(A)-Q(A)|}\label{eq:}
\end{equation}
where $\mathcal{F}$ is the $\sigma$-algebra of the probability space.

In this paper, some achievability schemes involves a random codebook
$\mathcal{C}$ (or a random binning $\mathcal{B}$). For simplicity,
we also denote the induced conditional distribution $P_{X|\mathcal{C}=c}$
(given $\mathcal{C}=c$) as $P_{X}$ (suppressing the condition $\mathcal{C}=c$),
which can be seen as a\emph{ random pmf}.

For any pmfs $P_{X}$ and $Q_{X}$ on $\mx$, we write $P_{X}\stackrel{\epsilon}{\approx}Q_{X}$
if $\tv{P_{X}-Q_{X}}<\epsilon$ for non-random pmfs, or $\e_{\mathcal{C}}\tv{P_{X}-Q_{X}}<\epsilon$
for random pmfs. For any two sequences of pmfs $P_{X^{(n)}}$ and
$Q_{X^{(n)}}$ on $\mx^{(n)}$ (where $\mx^{(n)}$ is arbitrary and
it differs from $\mx^{n}$ which is a Cartesian product), we write
$P_{X^{(n)}}\stackrel{}{\approx}Q_{X^{(n)}}$ if $\lim_{n\rightarrow\infty}\tv{P_{X^{(n)}}-Q_{X^{(n)}}}=0$
for non-random pmfs, or $\lim_{n\rightarrow\infty}\e_{\mathcal{C}}\tv{P_{X^{(n)}}-Q_{X^{(n)}}}=0$
for random pmfs.

The following properties of total variation distance hold.

\begin{property} \label{pr:properties} \cite{Schieler,Yassaee}
\textit{Total variation distance satisfies}:
\begin{enumerate}
\item \textit{If the support of} $P$ \textit{and} $Q$ \textit{is a countable
set} $\mathcal{X}$, \textit{then}
\begin{equation}
{\displaystyle \Vert P-Q\Vert_{TV}=\frac{1}{2}\sum_{x\in\mathcal{X}}|P(x)-Q(x)|}.
\end{equation}
\item \textit{Let} $\epsilon>0$ \textit{and let} $f(x)$ \textit{be a function
with bounded range of width} $b>0$. \textit{Then}
\begin{equation}
P_{X}\stackrel{\epsilon}{\approx}Q_{X}\Rightarrow|\mathbb{E}_{P}f(X)-\mathbb{E}_{Q}f(X)|<\epsilon b,\label{eq:-6}
\end{equation}
\textit{where} $\mathbb{E}_{P}$ \textit{indicates that the expectation
is taken with respect to the distribution} $P.$
\item $P_{X^{(n)}}\apx{}Q_{X^{(n)}}\Rightarrow P_{X^{(n)}}P_{Y^{(n)}|X^{(n)}}\apx{}Q_{X^{(n)}}P_{Y^{(n)}|X^{(n)}}$,\\
$P_{X^{(n)}}P_{Y^{(n)}|X^{(n)}}\apx{}Q_{X^{(n)}}Q_{Y^{(n)}|X^{(n)}}\Rightarrow P_{X^{(n)}}\apx{}Q_{X^{(n)}}$.
\item For any two sequences of non-random pmfs $P_{X^{(n)}Y^{(n)}}$ and
$Q_{X^{(n)}Y^{(n)}}$, if $P_{X^{(n)}}P_{Y^{(n)}|X^{(n)}}\stackrel{}{\approx}Q_{X^{(n)}}Q_{Y^{(n)}|X^{(n)}}$,
then there exists a sequence $x^{(n)}\in\mx^{(n)}$ such that $P_{Y^{(n)}|X^{(n)}=x^{(n)}}\stackrel{}{\approx}Q_{Y^{(n)}|X^{(n)}=x^{(n)}}$.
\item If $P_{X^{(n)}}\apx{}Q_{X^{(n)}}$ and $P_{X^{(n)}}P_{{Y^{(n)}}|{X^{(n)}}}\apx{}P_{X^{(n)}}Q_{{Y^{(n)}}|{X^{(n)}}}$,
then $P_{{X^{(n)}}}P_{{Y^{(n)}}|{X^{(n)}}}\apx{}Q_{{X^{(n)}}}Q_{{Y^{(n)}}|{X^{(n)}}}$.
\end{enumerate}
\end{property}

\section{(Conditional) Maximal Correlation}

In this section, we first define several correlations, including (Pearson)
correlation, correlation ratio, and maximal correlation, and then
study their properties. These concepts and properties will be used
to define and investigate information-correlation functions in subsequent
sections.

In this section, we assume all alphabets are general (not limited
to finite or countable) unless otherwise stated.

\subsection{\textit{\emph{Definition}}}
\begin{defn}
For any random variables $X$ and $Y$ with alphabets $\mathcal{X}\subseteq\mathbb{R}$
and $\mathcal{Y}\subseteq\mathbb{R}$, the (Pearson) correlation of
$X$ and $Y$ is defined by
\begin{equation}
\rho(X;Y)=\left\{ \begin{array}{ll}
\frac{\textrm{cov}(X,Y)}{\sqrt{\textrm{var}(X)}\sqrt{\textrm{var}(Y)}}, & \textrm{i}\textrm{f }\textrm{var}(X)\textrm{var}(Y)>0,\\
0, & \textrm{i}\textrm{f }\textrm{var}(X)\textrm{var}(Y)=0.
\end{array}\right.
\end{equation}
Moreover, the conditional correlation of $X$ and $Y$ given another
random variable $U$ is defined by
\begin{equation}
\rho(X;Y|U)=\left\{ \begin{array}{ll}
\frac{\mathbb{E}[\textrm{cov}(X,Y|U)]}{\sqrt{\mathbb{E}[\textrm{var}(X|U)]}\sqrt{\mathbb{E}[\textrm{var}(Y|U)]}}, & \textrm{i}\textrm{f }\mathbb{E}[\textrm{var}(X|U)]\mathbb{E}[\textrm{var}(Y|U)]>0,\\
0, & \textrm{i}\textrm{f }\mathbb{E}[\textrm{var}(X|U)]\mathbb{E}[\textrm{var}(Y|U)]=0.
\end{array}\right.
\end{equation}
\end{defn}
\begin{defn}
For any random variables $X$ and $Y$ with alphabets $\mathcal{X}\subseteq\mathbb{R}$
and $\mathcal{Y}$, the correlation ratio of $X$ on $Y$ is defined
by
\begin{equation}
{\displaystyle \theta(X;Y)=\sup_{g}\rho(X;g(Y))},
\end{equation}
where the supremum is taken over all the functions $g:\mathcal{Y}\mapsto\mathbb{R}$.
Moreover, the conditional correlation ratio of $X$ on $Y$ given
another random variable $U$ with alphabet $\mathcal{U}$ is defined
by
\begin{equation}
{\displaystyle \theta(X;Y|U)=\sup_{g}\rho(X;g(Y,U)|U)},
\end{equation}
where the supremum is taken over all the functions $g:\mathcal{Y}\times\mathcal{U}\mapsto\mathbb{R}.$
\end{defn}
\begin{rem}
Note that in general $\theta(X;Y)\neq\theta(Y;X)$ and $\theta(X;Y|U)\neq\theta(Y;X|U)$.
\end{rem}
\begin{defn}
\label{def:For-any-random}For any random variables $X$ and $Y$
with alphabets $\mathcal{X}$ and $\mathcal{Y}$, the maximal correlation
of $X$ and $Y$ is defined by
\begin{equation}
{\displaystyle \rho_{m}(X;Y)=\sup_{f,g}\rho(f(X);g(Y))},\label{eq:-1}
\end{equation}
where the supremum is taken over all the functions $f:\mathcal{X}\mapsto\mathbb{R},g:\mathcal{Y}\mapsto\mathbb{R}$.
Moreover, the conditional maximal correlation of $X$ and $Y$ given
another random variable $U$ with alphabet $\mathcal{U}$ is defined
by
\begin{equation}
{\displaystyle \rho_{m}(X;Y|U)=\sup_{f,g}\rho(f(X,U);g(Y,U)|U)},\label{eq:-2}
\end{equation}
where the supremum is taken over all the functions $f:\mathcal{X}\times\mathcal{U}\mapsto\mathbb{R},g:\mathcal{Y}\times\mathcal{U}\mapsto\mathbb{R}$.
\end{defn}
It is easy to verify that
\begin{equation}
{\displaystyle \rho_{m}(X;Y|U)=\sup_{f}\theta(f(X,U);Y|U)}.
\end{equation}

Note that the unconditional versions of correlation coefficient, correlation
ratio, and maximal correlation have been  well studied in literature.
The conditional versions are first introduced by Beigi and Gohari
recently \cite{Beigi}, where it is named as \emph{maximal correlation
of a box} and used to study the problem of non-local correlations.
In this paper, we will well study conditional maximal correlation
(and conditional correlation ratio), and give some useful properties.

\subsection{Properties }

According to the definition, maximal correlation remains the same
after applying bijective transform (one-to-one correspondence) on
each of the variables. Hence it is robust to bijective transform.
Furthermore, for finite valued random variables maximal correlation
$\rho_{m}(X;Y|U)$ can be characterized by the second largest singular
value $\lambda_{2}(u)$ of the matrix $Q_{u}$ with entries
\begin{equation}
Q_{u}(x,y):={\displaystyle \frac{p(x,y|u)}{\sqrt{p(x|u)p(y|u)}}=\frac{p(x,y,u)}{\sqrt{p(x,u)p(y,u)}}.}
\end{equation}
\begin{lem}
\label{lem:Singular-value-characterization}(Singular value characterization).
For any random variables $X,Y,U,$
\begin{equation}
{\displaystyle \rho_{m}(X;Y|U)=\sup_{u:P(u)>0}\lambda_{2}(u)}.\label{eq:-3}
\end{equation}
\end{lem}
\begin{rem}
This shows the conditional maximal correlation is consistent with
the unconditional version $(U=\emptyset)$ \cite{Witsenhausen}
\begin{equation}
\rho_{m}(X;Y)=\lambda_{2}.
\end{equation}
Furthermore, for any random variables $X,Y,U$ with finite alphabets,
 the supremum in \eqref{eq:-1}, \eqref{eq:-2} and \eqref{eq:-3}
is actually a maximum.
\end{rem}
The proof of this lemma is given in Appendix \ref{sec:Proof-of-Lemma-sing}.
This lemma gives a simple approach to compute (conditional) maximal
correlation. Observe that $\lambda_{2}(u)$ is equal to the maximal
correlation $\rho_{m}(X;Y|U=u)$ between $X$ and $Y$ under condition
$U=u$, and under distribution $P_{XY|U=u}$. Hence Lemma \ref{lem:Singular-value-characterization}
leads to the following result.
\begin{lem}
\label{lem:Alternative-characterization}(Alternative characterization).
For any random variables $X,Y,U,$
\begin{equation}
{\displaystyle \rho_{m}(X;Y|U)=\sup_{u:P(u)>0}\rho_{m}(X;Y|U=u)}.\label{eq:-4}
\end{equation}
\end{lem}
Note that the right-hand side of \eqref{eq:-4} was first defined
by Beigi and Gohari \cite{Beigi}. This lemma implies the equivalence
between the conditional maximal correlation defined by us and that
defined by Beigi and Gohari.

Furthermore, Lemmas \ref{lem:Singular-value-characterization} and
\ref{lem:Alternative-characterization} also hold for continuous random
variables, if the constraint of $P(u)>0$ is replaced with $p(u)>0$.
Here $p(u)$ denotes the probability density function (pdf) of $U$.
Notice that Lemmas \ref{lem:Singular-value-characterization} and
\ref{lem:Alternative-characterization} imply that $\rho_{m}(X;Y|U)$
can be different for different distributions of $X,Y$, even if the
distributions are only different  up to a zero measure set. In measure
theory, people usually do not care the difference with zero measure.
Therefore, we refine the definition of conditional maximal correlation
for continuous random variables by defining a robust version as follows.

\begin{equation}
{\displaystyle \widetilde{\rho}_{m}(X;Y|U):=\inf_{q_{XYU}:q_{XYU}=p_{XYU}\textrm{ a.s.}}\rho_{m,q}(X;Y|U)},\label{eq:-74}
\end{equation}
for continuous random variables $X,Y,U,$ with pdf $p_{XYU}$. We
name $\widetilde{\rho}_{m}(X;Y|U)$ as \emph{robust conditional maximal
correlation}. Obviously, for discrete random variables case, robust
conditional maximal correlation is consistent with conditional maximal
correlation. Moreover, if we take $\inf_{q_{XYU}:q_{XYU}=p_{XYU}\textrm{ a.s.}}$
operation on each side of an equality or inequality about $q_{XYU}$,
it usually does not change the equality or inequality. Hence in this
paper, we only consider conditional maximal correlations rather than
their robust versions.
\begin{lem}
\label{lem:TV-bound-on}(TV bound on maximal correlation). For any
random variables $X,Y,U$ with finite alphabets,
\begin{equation}
\rho_{m,Q}(X;Y|U)\geq\frac{\rho_{m,P}(X;Y|U)-\frac{4\delta}{P_{m}}}{1+\frac{4\delta}{P_{m}}},\label{eq:-7}
\end{equation}
where $P_{m}={\displaystyle \min_{x,y,u:P(x,y,u)>0}P(x,y|u)}$, and
${\displaystyle \delta=\max_{u:P(u)>0}\Vert P_{XY|U=u}-Q_{XY|U=u}\Vert_{TV}.}$
\end{lem}
\begin{rem}
Lemma \ref{lem:TV-bound-on} implies
\begin{equation}
{\displaystyle \frac{\rho_{m,P}(X;Y|U)-\frac{4\delta}{P_{m}}}{1+\frac{4\delta}{P_{m}}}\leq\rho_{m,Q}(X;Y|U)\leq\left(1+\frac{4\delta}{Q_{m}}\right)\rho_{m,P}(X;Y|U)+\frac{4\delta}{Q_{m}}},\label{eq:-56}
\end{equation}
where $Q_{m}={\displaystyle \min_{x,y,u:Q(x,y,u)>0}Q(x,y|u)}$.
\end{rem}
\begin{IEEEproof}
Assume $u$ achieves the supremum in \eqref{eq:-4}, and $f,g$ satisfying
$\mathbb{E}_{P}[f(X,U)|U=u]=0,\mathbb{E}_{P}[g(Y,U)|U=u]=0,\textrm{v}\textrm{a}\textrm{r}_{P}[f(X,U)|U=u]=1,\textrm{v}\textrm{a}\textrm{r}_{P}[g(Y,U)|U=u]=1$,
achieves $\rho_{m,P}(X;Y|U)$. Then $P(x|u)f^{2}(x,u)\leq\sum_{x}P(x|u)f^{2}(x,u)=1$
for any $x,u$, i.e.,
\begin{equation}
|f(x,{\displaystyle u)|\leq\frac{1}{\sqrt{P(x|u)}}}\label{eq:-5}
\end{equation}
for any $x,u$ such that $P(x|u)>0$. Furthermore, for any $x,u$
such that $P(x|u)>0,$ we have $P(x|u)\geq P(x,y|u)\geq P_{m}$. Hence
\eqref{eq:-5} implies
\begin{equation}
|f(x,{\displaystyle u)|\leq\frac{1}{\sqrt{P_{m}}}}.
\end{equation}
Similarly, we have
\begin{equation}
|g(y,{\displaystyle u)|\leq\frac{1}{\sqrt{P_{m}}}}.
\end{equation}
According to Property \eqref{eq:-6}, the following inequalities hold.
\begin{equation}
|{\displaystyle \mathbb{E}_{Q}[f(X,U)g(Y,U)|U]-\mathbb{E}_{P}[f(X,U)g(Y,U)|U]|\leq\frac{2\delta}{P_{m}}}
\end{equation}
\begin{equation}
|{\displaystyle \mathbb{E}_{Q}[f(X,U)|U]|\leq\frac{2}{\sqrt{P_{m}}}\Vert P_{X|U}-Q_{X|U}\Vert_{TV}\leq\frac{2\delta}{\sqrt{P_{m}}}}
\end{equation}
\begin{equation}
|{\displaystyle \mathbb{E}_{Q}[g(Y,U)|U]|\leq\frac{2}{\sqrt{P_{m}}}\Vert P_{Y|U}-Q_{Y|U}\Vert_{TV}\leq\frac{2\delta}{\sqrt{P_{m}}}}
\end{equation}
\begin{equation}
|{\displaystyle \mathbb{E}_{Q}[f^{2}(X,U)|U]-1|\leq\frac{2\delta}{P_{m}}}
\end{equation}
and
\begin{equation}
|{\displaystyle \mathbb{E}_{Q}[g^{2}(Y,U)|U]-1|\leq\frac{2\delta}{P_{m}}}.
\end{equation}
Therefore, we have
\begin{align}
{\displaystyle \rho_{m,Q}(X;Y|U=u)} & \geq\frac{\mathbb{E}_{Q}[f(X,U)g(Y,U)|U]-\mathbb{E}_{Q}[f(X,U)|U]\mathbb{E}_{Q}[g(Y,U)|U]}{\sqrt{\mathbb{E}_{Q}[f^{2}(X,U)|U]-\mathbb{E}_{Q}^{2}[f(X,U)|U]}\sqrt{\mathbb{E}_{Q}[g^{2}(Y,U)|U]-\mathbb{E}_{Q}^{2}[g(Y,U)|U]}}\\
 & {\displaystyle \geq\frac{\mathbb{E}_{Q}[f(X,U)g(Y,U)|U]-\frac{4\delta}{P_{m}}}{\sqrt{\mathbb{E}_{Q}[f^{2}(X,U)|U]-\frac{4\delta}{P_{m}}}\sqrt{\mathbb{E}_{Q}[g^{2}(Y,U)|U]-\frac{4\delta}{P_{m}}}}}\\
 & {\displaystyle \geq\frac{\rho_{m,P}(X;Y|U=u)-\frac{4\delta}{P_{m}}}{\sqrt{1+\frac{4\delta}{P_{m}}}\sqrt{1+\frac{4\delta}{P_{m}}}}}\\
 & ={\displaystyle \frac{\rho_{m,P}(X;Y|U=u)-\frac{4\delta}{P_{m}}}{1+\frac{4\delta}{P_{m}}}}.
\end{align}
\end{IEEEproof}
\begin{lem}
(Continuity and discontinuity). Assume $X,Y,U$ have finite alphabets.
Then given $P_{U},$ $\rho_{m}(X;Y|U)$ is continuous in $P_{XY|U}$.
Given $P_{XY|U},$ $\rho_{m}(X;Y|U)$ is continuous on $\{P_{U}:P_{U}(u)>0,\forall u\in\mathcal{U}\}.$
But in general, $\rho_{m}(X;Y|U)$ is discontinuous in $P_{XYU}$.
\end{lem}
\begin{IEEEproof}
\eqref{eq:-56} implies for given $P_{U}$, as ${\displaystyle \max_{u:P(u)>0}\Vert P_{XY|U=u}-Q_{XY|U=u}\Vert_{TV}\rightarrow0,}$
$\rho_{m,Q}(X;Y|U)\rightarrow\rho_{m,P}(X;Y|U)$. Hence for given
$P_{U},$ $\rho_{m,P}(X;Y|U)$ is continuous in $P_{XY|U}$. Furthermore,
since given $P_{XY|U},$ $\rho_{m}(X;Y|U)={\displaystyle \sup_{u:P(u)>0}\lambda_{2}(u)}$,
we have for given $P_{XY|U},$ $\rho_{m}(X;Y|U)$ is continuous on
$\{P_{U}:P_{U}(u)>0,\forall u\in\mathcal{U}\}.$ But it is worth noting
that $\rho_{m}(X;Y|U)$ may be discontinuous at $P_{U}$ such that
$P_{U}(u)=0$ for some $u\in\mathcal{U}$. Therefore, $Q_{XYU}\rightarrow P_{XYU}$
in total variation sense does not necessarily imply $\rho_{m,Q}(X;Y|U)\rightarrow\rho_{m}(X;Y|U)$.
That is, the conditional maximal correlation may be discontinuous
in probability distribution $P_{XYU}$.
\end{IEEEproof}
Furthermore, some other properties hold.
\begin{lem}
(Concavity). Given $P_{XY|U},$ $\rho_{m}(X;Y|U)$ is concave in
$P_{U}.$
\end{lem}
\begin{IEEEproof}

Fix $P_{XY|U}$. Assume $R_{U}=\lambda P_{U}+\left(1-\lambda\right)Q_{U}$,
$\lambda\in(0,1)$, then by Lemma \ref{lem:Alternative-characterization},
we have
\begin{align}
\rho_{m,R}(X;Y|U) & =\sup_{u:R(u)>0}\rho_{m}(X;Y|U=u)\label{eq:-4-1}\\
 & =\sup_{u:P(u)>0\textrm{ or }Q(u)>0}\rho_{m}(X;Y|U=u)\\
 & =\max\left\{ \sup_{u:P(u)>0}\rho_{m}(X;Y|U=u),\sup_{u:Q(u)>0}\rho_{m}(X;Y|U=u)\right\} \\
 & =\max\left\{ \rho_{m,P}(X;Y|U),\rho_{m,Q}(X;Y|U)\right\} .
\end{align}
Hence $\rho_{m,R}(X;Y|U)\geq\lambda\rho_{m,P}(X;Y|U)+\left(1-\lambda\right)\rho_{m,Q}(X;Y|U)$,
i.e., $\rho_{m}(X;Y|U)$ is concave in $P_{U}.$
\end{IEEEproof}
\begin{lem}
\label{lem:relationship}For any random variables $X,Y,Z,U$, the
following inequalities hold.
\begin{equation}
0\leq|\rho(X;Y|U)|\leq\theta(X;Y|U)\leq\rho_{m}(X;Y|U)\leq1.
\end{equation}
Moreover, $\rho_{m}(X;Y|U)=0$ if and only if $X$ and $Y$ are conditionally
independent given $U;$ $\rho_{m}(X;Y|U)=1$ if and only if $X$ and
$Y$ have Gács-Körner common information given $U.$
\end{lem}
\begin{IEEEproof}
\begin{align}
|\mathbb{E}[\textrm{cov}(X,Y|U)]| & =|\mathbb{E}[(X-\mathbb{E}[X|U])(Y-\mathbb{E}[Y|U])]|\\
 & \leq\sqrt{\mathbb{E}[(X-\mathbb{E}[X|U])^{2}]\mathbb{E}[(Y-\mathbb{E}[Y|U])^{2}]}\label{eq:-8}\\
 & =\sqrt{\mathbb{E}[\textrm{v}\textrm{a}\textrm{r}(X|U)]\mathbb{E}[\textrm{v}\textrm{a}\textrm{r}(Y|U)]},
\end{align}
where \eqref{eq:-8} follows from the Cauchy-Schwarz inequality. Hence
\begin{equation}
0\leq|\rho(X;Y|U)|\leq1
\end{equation}
which further implies
\begin{equation}
0\leq|\rho(X;Y|U)|\leq\theta(X;Y|U)\leq\rho_{m}(X;Y|U)\leq1
\end{equation}
since both $\theta(X;Y|U)$ and $\rho_{m}(X;Y|U)$ are conditional
correlations for some variables.

If $X$ and $Y$ are conditionally independent given $U$, then for
any functions $f$ and $g$, $f(X,U)$ and $g(Y,U)$ are also conditionally
independent given $U$. This leads to $\rho_{m}(X;Y|U)=0.$

Conversely, if $\rho_{m}(X;Y|U)=0$, then
\begin{equation}
\rho(f(X,U);g(Y,U)|U)=0\label{eq:-9}
\end{equation}
for any functions $f$ and $g$. For any $x,u$, set $f(X,U)=1\{X=x,U=u\}$
and $g(Y,U)=1\{Y=y,U=u\}$, then

\begin{align}
\mathbb{E}[\textrm{cov}(X,Y|U)] & =\mathbb{P}(X=x,Y=y|U=u)-\mathbb{P}(X=x|U=u)\mathbb{P}(Y=y|U=u)\\
 & =P_{XY|U}(x,y|u)-P_{X|U}(x|u)P_{Y|U}(y|u).
\end{align}
Hence \eqref{eq:-9} implies
\begin{equation}
P_{XY|U}(x,y|u)=P_{X|U}(x|u)P_{Y|U}(y|u).
\end{equation}
This implies $X$ and $Y$ are conditionally independent given $U$.
Therefore, $\rho_{m}(X;Y|U)=0$ if and only if $X$ and $Y$ are conditionally
independent given $U.$

Assume $X$ and $Y$ have Gács-Körner common information given $U$,
i.e., $f(X,U)=g(Y,U)$ with probability 1 for some functions $f$
and $g$ such that $H(f(X,U)|U)>0$. Then $\mathbb{E}\textrm{var}(f(X,U)|U)\mathbb{E}\textrm{var}(g(Y,U)|U)>0$,
and
\begin{equation}
\rho_{m}(X;Y|U)\geq\rho(f(X,U);g(Y,U)|U)\geq1.
\end{equation}
Combining this with $\rho_{m}(X;Y|U)\leq1$, we have $\rho_{m}(X;Y|U)=1.$

Assume $\rho_{m}(X;Y|U)=1$, then $f(X,U)=g(Y,U)$ with probability
1 for some functions $f$ and $g$ such that $\mathbb{E}\textrm{var}(f(X,U)|U)\mathbb{E}\textrm{var}(g(Y,U)|U)>0$,
or equivalently, $H(f(X,U)|U)>0$. This implies $X$ and $Y$ have
Gács-Körner common information given $U$. Therefore, $\rho_{m}(X;Y|U)=1$
if and only if $X$ and $Y$ have Gács-Körner common information given
$U$.
\end{IEEEproof}
\begin{lem}
\label{lem:For-any-random}For any random variables $X,Y,Z,U$, the
following properties hold.
\begin{equation}
\theta(X;YZ|U)\geq\theta(X;Y|U);\label{eq:-10}
\end{equation}
\begin{equation}
\rho_{m}(X;YZ|U)\geq\rho_{m}(X;Y|U);\label{eq:-11}
\end{equation}
\begin{align}
\theta(X;Y|U) & =\sqrt{\frac{\mathbb{E}[\textrm{var}(\mathbb{E}[X|YU]|U)]}{\mathbb{E}[\textrm{var}(X|U)]}}\nonumber \\
 & =\sqrt{1-\frac{\mathbb{E}[\textrm{var}(X|YU)]}{\mathbb{E}[\textrm{var}(X|U)]}};\label{eq:-12}
\end{align}
\begin{align}
\rho_{m}(X;Y|U) & =\sup_{f}{\displaystyle \sqrt{\frac{\mathbb{E}[\textrm{var}(\mathbb{E}[f(X,U)|YU]|U)]}{\mathbb{E}[\textrm{var}(f(X,U)|U)]}}}\nonumber \\
 & =\sup_{f}{\displaystyle \sqrt{1-\frac{\mathbb{E}[\textrm{var}(f(X,U)|YU)]}{\mathbb{E}[\textrm{var}(f(X,U)|U)]}}}.\label{eq:-13}
\end{align}
In particular if $U$ is degenerate, then the inequalities above reduce
to
\begin{equation}
\theta(X;YZ)\geq\theta(X;Y);
\end{equation}
\begin{equation}
\rho_{m}(X;YZ)\geq\rho_{m}(X;Y);
\end{equation}
\begin{align}
\theta(X;Y) & =\sqrt{\frac{\textrm{var}(\mathbb{E}[X|Y])}{\textrm{var}(X)}}\nonumber \\
 & =\sqrt{1-\frac{\mathbb{E}[\textrm{var}(X|Y)]}{\textrm{var}(X)}};
\end{align}
\begin{align}
\rho_{m}(X;Y) & =\sup_{f}{\displaystyle \sqrt{\frac{\textrm{var}(\mathbb{E}[f(X)|Y])}{\textrm{var}(f(X))}}}\nonumber \\
 & =\sup_{f}\sqrt{1-\frac{\mathbb{E}[\textrm{var}(f(X)|Y)]}{\textrm{var}(f(X))}}.
\end{align}
\end{lem}
\begin{rem}
Correlation ratio is also closely related to Minimum Mean Square Error
(MMSE). The optimal MMSE estimator is $\mathbb{E}[X|YU]$, hence the
variance of the MMSE for estimating $X$ given $(Y,U)$ is $\textrm{mmse}(X|YU)=\mathbb{E}(X-\mathbb{E}[X|YU])^{2}=\mathbb{E}[\textrm{var}(X|YU)]=\mathbb{E}[\textrm{var}(X|U)](1-\theta^{2}(X;Y|U)).$
\end{rem}
\begin{IEEEproof}
According to definitions of conditional correlation ratio and conditional
maximal correlation, \eqref{eq:-10} and \eqref{eq:-11} can be proven
easily.

In fact, we may, without loss of the generality, consider only such
function $g$ for which $\mathbb{E}[g(Y,U)|U=u]=0,\forall u$ and
$\textrm{var}(g(Y,U)|U=u)=1,\forall u$ and suppose $\mathbb{E}[X]=0,\mathbb{E}[\textrm{var}(X|U)]=1;$
for this case we have by the Cauchy-Schwarz inequality
\begin{align}
\mathbb{E}[\textrm{cov}(X,g(Y,U)|U)] & =\mathbb{E}[(X-\mathbb{E}[X|U])(g(Y,U)-\mathbb{E}[g(Y,U)|U])]\\
 & =\mathbb{E}[\mathbb{E}[(X-\mathbb{E}[X|U])|YU](g(Y,U)-\mathbb{E}[g(Y,U)|U])]\\
 & =\mathbb{E}[(\mathbb{E}[X|YU]-\mathbb{E}[X|U])(g(Y,U)-\mathbb{E}[g(Y,U)|U])]\\
 & \leq\sqrt{\mathbb{E}[(\mathbb{E}[X|YU]-\mathbb{E}[X|U])^{2}]\mathbb{E}[(g(Y,U)-\mathbb{E}[g(Y,U)|U])^{2}]}\\
 & =\sqrt{\mathbb{E}[\textrm{v}\textrm{a}\textrm{r}(\mathbb{E}[X|YU]|U)]\mathbb{E}[\textrm{v}\textrm{a}\textrm{r}(g(Y,U)|U)]}.
\end{align}

Therefore,
\begin{align}
{\displaystyle \theta(X;Y|U)} & =\sup_{g}\frac{\mathbb{E}[\textrm{cov}(X,g(Y,U)|U)]}{\sqrt{\mathbb{E}[\textrm{v}\textrm{a}\textrm{r}(X|U)]\mathbb{E}[\textrm{v}\textrm{a}\textrm{r}(g(Y,U)|U)]}}\\
 & \leq{\displaystyle \sqrt{\frac{\mathbb{E}[\textrm{v}\textrm{a}\textrm{r}(\mathbb{E}[X|YU]|U)]}{\mathbb{E}[\textrm{v}\textrm{a}\textrm{r}(X|U)]}}}.
\end{align}
It is easy to verify that equality holds if and only if $g(Y,U)=\alpha\mathbb{E}[X|YU]$
for some constant $\alpha>0$. Hence
\begin{equation}
\theta(X;Y|U)=\sqrt{\frac{\mathbb{E}[\textrm{v}\textrm{a}\textrm{r}(\mathbb{E}[X|YU]|U)]}{\mathbb{E}[\textrm{v}\textrm{a}\textrm{r}(X|U)]}}.
\end{equation}

Furthermore, by law of total variance
\begin{equation}
\textrm{var}(Y)=\mathbb{E}\textrm{var}(Y|X)+\textrm{v}\textrm{a}\textrm{r}(\mathbb{E}(Y|X))
\end{equation}
and the conditional version
\begin{equation}
\mathbb{E}[\textrm{var}(X|U)]=\mathbb{E}[\textrm{var}(X|YU)]+\mathbb{E}[\textrm{var}(\mathbb{E}[X|YU]|U)],
\end{equation}
we have
\begin{align}
\theta(X;Y|U) & =\sqrt{\frac{\mathbb{E}[\textrm{v}\textrm{a}\textrm{r}(\mathbb{E}[X|YU]|U)]}{\mathbb{E}[\textrm{v}\textrm{a}\textrm{r}(X|U)]}}\nonumber \\
 & ={\displaystyle \sqrt{1-\frac{\mathbb{E}[\textrm{v}\textrm{a}\textrm{r}(X|YU)]}{\mathbb{E}[\textrm{v}\textrm{a}\textrm{r}(X|U)]}}}.\label{eq:-14}
\end{align}

Furthermore, since ${\displaystyle \rho_{m}(X;Y|U)=\sup_{f}\theta(f(X,U);Y|U),}$
\eqref{eq:-13} follows straightforwardly from \eqref{eq:-14}.
\end{IEEEproof}
\begin{lem}
\label{lem:Correlation-ratio-equality}(Correlation ratio equality).
For any random variables $X,Y,U,$
\begin{align}
1-\theta^{2}(X;YZ|U) & =(1-\theta^{2}(X;Z|U))(1-\theta^{2}(X;Y|ZU));\label{eq:-15}\\
1-\rho_{m}^{2}(X;YZ|U) & \geq(1-\rho_{m}^{2}(X;Z|U))(1-\rho_{m}^{2}(X;Y|ZU));\\
\theta(X;YZ|U) & \geq\theta(X;Y|ZU);\label{eq:-16}\\
\rho_{m}(X;YZ|U) & \geq\rho_{m}(X;Y|ZU).\label{eq:-51}
\end{align}
\end{lem}
\begin{rem}
 \eqref{eq:-51} is very similar to $I(X;YZ|U)\geq I(X;Y|ZU)$. Furthermore,
$\rho_{m}(X;Y|UV)\geq\rho_{m}(X;Y|U)$ or $\rho_{m}(X;Y|UV)\leq\rho_{m}(X;Y|U)$
does not always hold. This is also similar to that $I(X;Y|UV)\geq I(X;Y|U)$
or $I(X;Y|UV)\leq I(X;Y|U)$ does not always hold.
\end{rem}
\begin{IEEEproof}
From \eqref{eq:-12}, we have
\begin{equation}
1-{\displaystyle \theta^{2}(X;YZ|U)=\frac{\mathbb{E}[\textrm{v}\textrm{a}\textrm{r}(X|YZU)]}{\mathbb{E}[\textrm{v}\textrm{a}\textrm{r}(X|U)]}},
\end{equation}
\begin{equation}
1-{\displaystyle \theta^{2}(X;Z|U)=\frac{\mathbb{E}[\textrm{v}\textrm{a}\textrm{r}(X|ZU)]}{\mathbb{E}[\textrm{v}\textrm{a}\textrm{r}(X|U)]}},
\end{equation}
and
\begin{equation}
1-{\displaystyle \theta^{2}(X;Y|ZU)=\frac{\mathbb{E}[\textrm{v}\textrm{a}\textrm{r}(X|YZU)]}{\mathbb{E}[\textrm{v}\textrm{a}\textrm{r}(X|ZU)]}}.
\end{equation}
Hence \eqref{eq:-15} follows immediately.

Suppose $f$ achieves $\rho_{m}(X;YZ|U)$, i.e., the supremum in \eqref{eq:-2},
then
\begin{align}
1-\rho_{m}^{2}(X;YZ|U) & =1-\theta^{2}(f(X,U);YZ|U)\\
 & =(1-\theta^{2}(f(X,U);Z|U))(1-\theta^{2}(f(X,U);Y|ZU))\\
 & \geq(1-\rho_{m}^{2}(X;Z|U))(1-\rho_{m}^{2}(X;Y|ZU)).
\end{align}

Furthermore, $\theta^{2}(X;Z|U)\geq0$, hence \eqref{eq:-16} follows
immediately from \eqref{eq:-15}.

Suppose $f'$ achieves $\rho_{m}(X;Y|ZU)$, then
\begin{equation}
\rho_{m}(X;Y|ZU)=\theta(f'(X,U);Y|ZU)\leq\theta(f'(X,U);YZ|U)\leq\rho_{m}(X;YZ|U).
\end{equation}
\end{IEEEproof}
\begin{lem}
For any $P_{UXYV}$ such that $U\rightarrow X\rightarrow Y$ and $X\rightarrow Y\rightarrow V$,
we have
\begin{equation}
\rho_{m}(UX;VY)=\max\{\rho_{m}(X;Y),\rho_{m}(U;V|XY)\}.
\end{equation}
\end{lem}
\begin{rem}
A similar result can be found in \cite[Eqn. (4)]{Beigi}, where Beigi
and Gohari only proved the equality above as an inequality.
\end{rem}
\begin{IEEEproof}
Beigi and Gobari \cite[Eqn. (4)]{Beigi} have proven ${\displaystyle \rho_{m}(UX;VY)\leq\max\{\rho_{m}(X;Y),\rho_{m}(U;V|XY)\}}$.
Hence we only need to prove that ${\displaystyle \rho_{m}(UX;VY)\geq\max\{\rho_{m}(X;Y),\rho_{m}(U;V|XY)\}}$.
According to the definition, $\rho_{m}(UX;VY)\geq\rho_{m}(X;Y)$ is
straightforward. From \eqref{eq:-51} of Lemma \ref{lem:Correlation-ratio-equality},
we have $\rho_{m}(UX;VY)\geq\rho_{m}(UX;V|Y)\geq\rho_{m}(U;V|XY)$.
This completes the proof.
\end{IEEEproof}
We also prove that conditioning reduces covariance gap as shown in
the following lemma, the proof of which is given in Appendix \ref{sec:Proof-of-Lemma-cond}.
\begin{lem}
\label{lem:Conditioning-reduces-covariance}(Conditioning reduces
covariance gap). For any random variables $X,Y,Z,U,$
\begin{equation}
\sqrt{\mathbb{E}\textrm{var}(X|ZU)\mathbb{E}\textrm{var}(Y|ZU)}-\mathbb{E}\textrm{cov}(X,Y|ZU)\leq\sqrt{\mathbb{E}\textrm{var}(X|Z)\mathbb{E}\textrm{var}(Y|Z)}-\mathbb{E}\textrm{cov}(X,Y|Z),
\end{equation}
i. e.,
\begin{equation}
\sqrt{(1-\theta^{2}(X;U|Z))(1-\theta^{2}(Y;U|Z))}(1-\rho(X,Y|ZU))\leq1-\rho(X,Y|Z).
\end{equation}
In particular, if $Z$ is degenerate, then
\begin{equation}
\sqrt{\mathbb{E}\textrm{var}(X|U)\mathbb{E}\textrm{var}(Y|U)}-\mathbb{E}\textrm{cov}(X,Y|U)\leq\sqrt{\textrm{var}(X)\textrm{var}(Y)}-\textrm{cov}(X,Y),\label{eq:-46-1}
\end{equation}
i. e.,
\begin{equation}
\sqrt{(1-\theta^{2}(X;U))(1-\theta^{2}(Y;U))}(1-\rho(X,Y|U))\leq1-\rho(X,Y).
\end{equation}
\end{lem}
\begin{rem}
The following two inequalities follows immediately.
\begin{equation}
\sqrt{(1-\rho_{m}^{2}(X;U|Z))(1-\theta^{2}(Y;U|Z))}(1-\theta(X,Y|ZU))\leq1-\theta(X,Y|Z),
\end{equation}
and
\begin{equation}
\sqrt{(1-\rho_{m}^{2}(X;U|Z))(1-\rho_{m}^{2}(Y;U|Z))}(1-\rho_{m}(X,Y|ZU))\leq1-\rho_{m}(X,Y|Z).
\end{equation}
\end{rem}
Furthermore, there are also some other remarkable properties.
\begin{lem}
\label{lem:MCsequence}(Tensorization). Assume given $U,$ $(X^{n},Y^{n})$
is a sequence of pairs of conditionally independent random variables,
then we have
\begin{equation}
{\displaystyle \rho_{m}(X^{n};Y^{n}|U)=\sup_{1\leq i\leq n}\rho_{m}(X_{i};Y_{i}|U)}.
\end{equation}
\end{lem}
\begin{IEEEproof}
The unconditional version
\begin{equation}
{\displaystyle \rho_{m}(X^{n};Y^{n})=\sup_{1\leq i\leq n}\rho_{m}(X_{i};Y_{i})},
\end{equation}
for a sequence of pairs of independent random variables $(X^{n},Y^{n})$
is proven in \cite[Thm. 1]{Witsenhausen}. Using this result and Lemma
\ref{lem:Singular-value-characterization}, we have

\begin{align}
{\displaystyle \rho_{m}(X^{n};Y^{n}|U)} & =\sup_{u:P(u)>0}\rho_{m}(X^{n};Y^{n}|U=u)\\
 & ={\displaystyle \sup_{u:P(u)>0}\sup_{1\leq i\leq n}\rho_{m}(X_{i};Y_{i}|U=u)}\\
 & ={\displaystyle \sup_{1\leq i\leq n}\sup_{u:P(u)>0}\rho_{m}(X_{i};Y_{i}|U=u)}\\
 & ={\displaystyle \sup_{1\leq i\leq n}\rho_{m}(X_{i};Y_{i}|U)}.
\end{align}
\end{IEEEproof}
\begin{lem}
(Gaussian case). For jointly Gaussian random variables $X,Y,U$, we
have
\begin{align}
 & \rho_{m}(X;Y)=\theta(X;Y)=\theta(Y;X)=|\rho(X;Y)|,\label{eq:-17}\\
 & \rho_{m}(X;Y|U)=\theta(X;Y|U)=\theta(Y;X|U)=|\rho(X;Y|U)|.\label{eq:-18}
\end{align}
\end{lem}
\begin{IEEEproof}
The unconditional version \eqref{eq:-17} is proven in \cite[Sec. IV, Lem. 10.2]{Rozanov}.
On the other hand, given $U=u,$ $(X,Y)$ also follows jointly Gaussian
distribution, and $\rho(X;Y|U=u)=\rho(X;Y|U)$ for different $u$.
Hence $\rho_{m}(X;Y|U)={\displaystyle \sup_{u:P(u)>0}\rho_{m}(X;Y|U=u)=\sup_{u:P(u)>0}|\rho(X;Y|U=u)|=|\rho(X;Y|U)|.}$

Furthermore, both $\theta(X;Y|U)$ and $\theta(Y;X|U)$ are between
$\rho_{m}(X;Y|U)$ and $|\rho(X;Y|U)|$. Hence \eqref{eq:-18} holds.
\end{IEEEproof}
\begin{lem}
(Data processing inequality). If random variables $X,Y,Z,U$ form
a Markov chain $X\rightarrow(Z,U)\rightarrow Y,$ then
\begin{align}
|\rho(X;Y|U)| & \leq\theta(X;Z|U)\theta(Y;Z|U),\label{eq:-19}\\
\theta(X;Y|U) & \leq\theta(X;Z|U)\rho_{m}(Y;Z|U),\label{eq:-23}\\
\rho_{m}(X;Y|U) & \leq\rho_{m}(X;Z|U)\rho_{m}(Y;Z|U).\label{eq:-20}
\end{align}
Moreover, the equalities hold in \eqref{eq:-19}-\eqref{eq:-20},
if $(X,Z,U)$ and $(Y,Z,U)$ have the same joint distribution. In
particular if $U$ is degenerate, then
\begin{align}
|\rho(X;Y)| & \leq\theta(X;Z)\theta(Y;Z),\label{eq:-19-1}\\
\theta(X;Y) & \leq\theta(X;Z)\rho_{m}(Y;Z),\\
\rho_{m}(X;Y) & \leq\rho_{m}(X;Z)\rho_{m}(Y;Z).\label{eq:-20-1}
\end{align}
\end{lem}
\begin{IEEEproof}
Consider that
\begin{align}
\mathbb{E}[\textrm{cov}(X,Y|U)] & =\mathbb{E}[(X-\mathbb{E}[X|U])(Y-\mathbb{E}[Y|U])]\\
 & =\mathbb{E}[\mathbb{E}[(X-\mathbb{E}[X|U])(Y-\mathbb{E}[Y|U])|ZU]]\\
 & =\mathbb{E}[\mathbb{E}[X-\mathbb{E}[X|U]|ZU]\mathbb{E}[Y-\mathbb{E}[Y|U]|ZU]]\label{eq:-21}\\
 & =\mathbb{E}[(\mathbb{E}[X|ZU]-\mathbb{E}[X|U])(\mathbb{E}[Y|ZU]-\mathbb{E}[Y|U])]\\
 & \leq\sqrt{\mathbb{E}[(\mathbb{E}[X|ZU]-\mathbb{E}[X|U])^{2}]\mathbb{E}[(\mathbb{E}[Y|ZU]-\mathbb{E}[Y|U])^{2}]}\label{eq:-22}\\
 & =\sqrt{\mathbb{E}[\textrm{v}\textrm{a}\textrm{r}(\mathbb{E}[X|ZU]|U)]\mathbb{E}[\textrm{v}\textrm{a}\textrm{r}(\mathbb{E}[Y|ZU]|U)]}
\end{align}
where \eqref{eq:-21} follows by conditional independence, and \eqref{eq:-22}
follows the Cauchy-Schwarz inequality. Hence
\begin{align}
|\rho(X;Y|U)| & =\frac{\mathbb{E}[\textrm{c}\textrm{o}\textrm{v}(X,Y|U)]}{\sqrt{\mathbb{E}[\textrm{v}\textrm{a}\textrm{r}(X|U)]}\sqrt{\mathbb{E}[\textrm{v}\textrm{a}\textrm{r}(Y|U)]}}\\
 & {\displaystyle \leq\sqrt{\frac{\mathbb{E}[\textrm{v}\textrm{a}\textrm{r}(\mathbb{E}[X|ZU]|U)]\mathbb{E}[\textrm{v}\textrm{a}\textrm{r}(\mathbb{E}[Y|ZU]|U)]}{\mathbb{E}[\textrm{v}\textrm{a}\textrm{r}(X|U)]\mathbb{E}[\textrm{v}\textrm{a}\textrm{r}(Y|U)]}}}\\
 & =\theta(X;Z|U)\theta(Y;Z|U).
\end{align}

It is easy to verify the equalities hold if $(X,Z,U)$ and $(Y,Z,U)$
have the same joint distribution.

Similarly, \eqref{eq:-23} and \eqref{eq:-20} can be proven as well.
\end{IEEEproof}
Furthermore, correlation ratio and maximal correlation are also related
to rate-distortion theory.
\begin{lem}
(Relationship to rate-distortion function) Let $R_{X|U}\left(D\right)$
denote the conditional rate distribution function for source $X$
given $U$ with quadratic distortion measure $d\left(x,\hat{x}\right)=\left(x-\hat{x}\right)^{2}$.
Then from rate-distortion theory, we have
\begin{align}
I(X;Y|U) & \geq R_{X|U}(\mathbb{E}[\textrm{var}(X|YU)])\\
 & =R_{X|U}(\mathbb{E}[\textrm{var}(X|U)](1-\theta^{2}(X;Y|U)))\\
 & \geq R_{X|U}(\mathbb{E}[\textrm{var}(X|U)](1-\rho^{2}(X;Y|U))).
\end{align}
From Shannon lower bound,
\begin{align}
I(X;Y|U) & \geq R_{X|U}(\mathbb{E}[\textrm{var}(X|YU)])\\
 & {\displaystyle \geq h(X|U)-\frac{1}{2}\log(2\pi e\mathbb{E}[\textrm{var}(X|U)](1-\theta^{2}(X;Y|U)))}.
\end{align}
If $(X,U)$ is jointly Gaussian, then
\begin{align}
I(X;Y|U) & {\displaystyle \geq\frac{1}{2}\log^{+}(\frac{1}{1-\theta^{2}(X;Y|U)})}\\
 & {\displaystyle \geq\frac{1}{2}\log^{+}(\frac{1}{1-\rho^{2}(X;Y|U)})}.
\end{align}
In particular if $U$ is degenerate, then
\begin{align}
I(X;Y) & \geq R_{X}(\mathbb{E}[\textrm{var}(X|Y)])\\
 & =R_{X}(\textrm{var}(X)(1-\theta^{2}(X;Y))).
\end{align}
From Shannon lower bound,
\begin{align}
I(X;Y) & \geq R_{X}(\mathbb{E}[\textrm{var}(X|Y)])\\
 & {\displaystyle \geq h(X)-\frac{1}{2}\log(2\pi e(1-\theta^{2}(X;Y}))).
\end{align}
If $X$ is Gaussian, then
\begin{align}
I(X;Y) & \geq\frac{1}{2}\log^{+}(\frac{1}{1-\theta^{2}(X;Y)})\\
 & {\displaystyle \geq\frac{1}{2}\log^{+}(\frac{1}{1-\rho^{2}(X;Y)})}.
\end{align}
\end{lem}
From the properties above, it can be observed that maximal correlation
or correlation ratio has many similar properties as those of mutual
information, such as invariance to one-to-one transform, chain rule
(correlation ratio equality), data processing inequality, etc. On
the other hand, maximal correlation or correlation ratio also has
some different properties, such as for a sequence of pairs of independent
random variables, the mutual information between them is the sum of
mutual information of all pairs of components (i.e., additivity);
while the maximal correlation is the maximum one of the maximal correlations
of all pairs of components (i.e., tensorization).

\subsection{Extension: Smooth Maximal Correlation}

Next we extend maximal correlation to smooth version. Analogous extensions
can be found in \cite{Liu} and \cite{Liu2}, where  Rényi divergence
and generalized Brascamp-Lieb-like (GBLL) rate are extended to the
corresponding smooth versions.
\begin{defn}
For any random variables $X$ and $Y$ with alphabets $\mathcal{X}\subseteq\mathbb{R}$
and $\mathcal{Y}\subseteq\mathbb{R}$, and $\epsilon\in\left(0,1\right)$,
the $\epsilon$-smooth (Pearson) correlation and the $\epsilon$-smooth
conditional (Pearson) correlation of $X$ and $Y$ given another random
variable $U$ are respectively defined by
\begin{equation}
{\displaystyle \widetilde{\rho}^{\epsilon}(X;Y):=\inf_{Q_{XY}:\left\Vert Q_{XY}-P_{XY}\right\Vert _{TV}\leq\epsilon}\rho_{Q}(X;Y)},\label{eq:-72-2-1}
\end{equation}
and
\begin{equation}
{\displaystyle \widetilde{\rho}^{\epsilon}(X;Y|U):=\inf_{Q_{XYU}:\left\Vert Q_{XYU}-P_{XYU}\right\Vert _{TV}\leq\epsilon}\rho_{Q}(X;Y|U)}.\label{eq:-72-2}
\end{equation}
\end{defn}
\begin{defn}
For any random variables $X$ and $Y$ with alphabets $\mathcal{X}\subseteq\mathbb{R}$
and $\mathcal{Y}$, and $\epsilon\in\left(0,1\right)$, the $\epsilon$-smooth
correlation ratio and the $\epsilon$-smooth conditional correlation
ratio of $X$ and $Y$ given another random variable $U$ are respectively
defined by
\begin{equation}
{\displaystyle \widetilde{\theta}^{\epsilon}(X;Y):=\inf_{Q_{XY}:\left\Vert Q_{XY}-P_{XY}\right\Vert _{TV}\leq\epsilon}\theta_{Q}(X;Y)},\label{eq:-72-2-1-1}
\end{equation}
and
\begin{equation}
{\displaystyle \widetilde{\theta}^{\epsilon}(X;Y|U):=\inf_{Q_{XYU}:\left\Vert Q_{XYU}-P_{XYU}\right\Vert _{TV}\leq\epsilon}\theta_{Q}(X;Y|U)}.\label{eq:-72-2-2}
\end{equation}
\end{defn}
\begin{defn}
\label{def:For-any-random-1}For any random variables $X$ and $Y$
with alphabets $\mathcal{X}$ and $\mathcal{Y}$, and $\epsilon\in\left(0,1\right)$,
the $\epsilon$-smooth maximal correlation and the $\epsilon$-smooth
conditional maximal correlation of $X$ and $Y$ given another random
variable $U$ are respectively defined by
\begin{equation}
{\displaystyle \widetilde{\rho}_{m}^{\epsilon}(X;Y):=\inf_{Q_{XY}:\left\Vert Q_{XY}-P_{XY}\right\Vert _{TV}\leq\epsilon}\rho_{m,Q}(X;Y)},\label{eq:-72-2-1-1-1}
\end{equation}
and
\begin{equation}
{\displaystyle \widetilde{\rho}_{m}^{\epsilon}(X;Y|U):=\inf_{Q_{XYU}:\left\Vert Q_{XYU}-P_{XYU}\right\Vert _{TV}\leq\epsilon}\rho_{m,Q}(X;Y|U)}.\label{eq:-81}
\end{equation}

According to definition, obviously we have
\begin{align}
\widetilde{\rho}^{\epsilon}(X;Y|U) & \leq\rho(X;Y|U),\\
\widetilde{\theta}^{\epsilon}(X;Y|U) & \leq\theta(X;Y|U),\\
\widetilde{\rho}_{m}^{\epsilon}(X;Y|U) & \leq\rho_{m}(X;Y|U).
\end{align}
Furthermore, note that adding $\inf_{Q_{XYU}:\left\Vert Q_{XYU}-P_{XYU}\right\Vert _{TV}\leq\epsilon}$
operation before both sides of an equality or inequality about $P_{XYU}$
does not change the equality or inequality. Hence some of above lemmas
still hold for $\epsilon$-smooth version, e.g., Lemmas \ref{lem:Singular-value-characterization},
\ref{lem:Alternative-characterization}, \ref{lem:relationship},
and \ref{lem:For-any-random}, and also \eqref{eq:-16} and \eqref{eq:-51}
of Lemma \ref{lem:Correlation-ratio-equality}.

\end{defn}

\section{Generalized Common Information: Information-Correlation Function}

In this section, we generalize the existing common informations, and
define $\beta$-approximate common information (or approximate information-correlation
function) and $\beta$-exact common information (or exact information-correlation
function), which measure how much information are approximately or
exactly $\beta$-correlated between two variables. Different from
the existing common informations, $\beta$-common information is a
function of conditional maximal correlation $\beta\in\left[0,1\right]$,
and hence it provides a soft-measure of common information.

As in the previous section, in this section we also assume all alphabets
are general unless otherwise stated.

\subsection{\textit{\emph{Definition}}}

Suppose $U$ is a common random variable extracted from $X,Y$, satisfying
privacy constraint $\rho_{m}(X;Y|U)\leq\beta$, then the $\beta$-private
information corresponding to $U$ should be $H(XY|U)$. We define
the $\beta$-private information as the maximum of such private informations
over all possible $U$.
\begin{defn}
For sources $X,Y$, and $\beta\in[0,1]$, the $\beta$-approximate
private information of $X$ and $Y$ is defined by
\begin{equation}
B_{\beta}(X;Y)={\displaystyle \sup_{P_{U|XY}:\rho_{m}(X;Y|U)\leq\beta}H(XY|U)}.
\end{equation}
\end{defn}
Common information is defined as $C_{\beta}(X;Y)=H(XY)-B_{\beta}(X;Y)$,
which is equivalent to the following definition.
\begin{defn}
For sources $X,Y$, and $\beta\in[0,1]$, the $\beta$-approximate
common information (or approximate information-correlation function)
of $X$ and $Y$ is defined by
\begin{equation}
C_{\beta}(X;Y)={\displaystyle \inf_{P_{U|XY}:\rho_{m}(X;Y|U)\leq\beta}I(XY;U)}.\label{eq:-24}
\end{equation}
\end{defn}
Similarly, exact common information can be generalized to $\beta$-exact
common information as well.
\begin{defn}
For sources $X,Y$, and $\beta\in[0,1]$, the $\beta$-exact common
information (rate) (or exact information-correlation function) of
$X$ and $Y$ is defined by
\begin{equation}
K_{\beta}(X;Y)={\displaystyle \lim_{n\rightarrow\infty}\inf_{P_{U_{n}|X^{n}Y^{n}}:\rho_{m}(X^{n};Y^{n}|U_{n})\leq\beta}\frac{1}{n}H(U_{n})}.
\end{equation}
\end{defn}
Furthermore, for $\beta\in(0,1]$, we also define
\begin{align}
C_{\beta^{-}}(X;Y) & =\lim_{\alpha\uparrow\beta}C_{\alpha}(X;Y),\\
K_{\beta^{-}}(X;Y) & =\lim_{\alpha\uparrow\beta}K_{\alpha}(X;Y).
\end{align}

\subsection{Properties }

\begin{figure}
\begin{centering}
\includegraphics[width=0.5\textwidth]{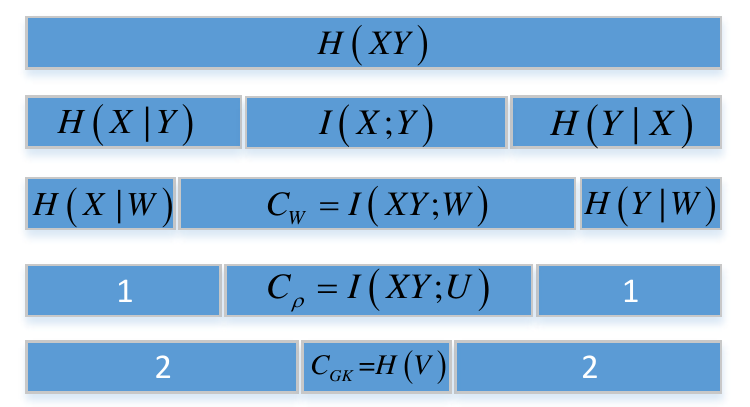}
\par\end{centering}
\caption{\label{fig:Illustration-of-the}Illustration of the relationship among
joint entropy, mutual information, Wyner common information, generalized
common information, and Gács-Körner common information, where $W,V$
and $U$ are the Wyner, Gács-Körner, and $\beta$-common random variables,
respectively, and Region 1 represents $H(XY|U)$ and Region 2 represents
$H(XY|V)$. These terms satisfy $C_{GK}\leq I(X;Y)\leq C_{W}$ and
$C_{GK}=C_{0}{\displaystyle \leq C_{\beta}\leq C_{1^{-}}(X;Y)=C_{W},\,\forall0\leq\beta<1}$.}
\end{figure}

These two generalized common informations have the following properties.
\begin{lem}
\label{lem:GCIproperties}(a) For the infimum in \eqref{eq:-24},
it suffices to consider the variable $U$ with alphabet $|\mathcal{U}|\leq|\mathcal{X}||\mathcal{Y}|+1.$

(b) For any random variables $X,Y,$ $C_{\beta}(X;Y)$ and $K_{\beta}(X;Y)$
are decreasing in $\beta.$ Moreover,
\begin{align}
 & C_{\beta}(X;Y)\leq K_{\beta}(X;Y),\textrm{for }0\leq\beta\leq1,\label{eq:-59}\\
 & C_{\beta}(X;Y)=K_{\beta}(X;Y)=0,\textrm{for }\rho_{m}(X;Y)\leq\beta\leq1,\label{eq:-60}\\
 & C_{0}(X;Y)=C_{W}(X;Y),\label{eq:-61}\\
 & K_{0}(X;Y)=K_{KLG}(X;Y),\label{eq:-62}\\
 & C_{1^{-}}(X;Y)=K_{1^{-}}(X;Y)=C_{GK}(X;Y),\label{eq:-63}
\end{align}
where $K_{KLG}(X;Y):=\lim_{n\rightarrow\infty}\inf_{P_{U_{n}|X^{n}Y^{n}}:X^{n}\rightarrow U_{n}\rightarrow Y^{n}}\frac{1}{n}H(U_{n})$
denotes the exact common information (rate) proposed by Kumar, Li,
and Gamal \cite{Kumar}.

(c) If $P_{U|X,Y}$ achieves the infimum in \eqref{eq:-24}, then
$\rho_{m}(X;Y|U)\leq\rho_{m}(X;Y|V)$ for any $V$ such that $XY\rightarrow U\rightarrow V$.
\end{lem}
\begin{rem}
For any random variables $X,Y,$ $C_{\beta}(X;Y)$ is decreasing in
$\beta$, but it is not necessarily convex or concave; see the Gaussian
source case in the next subsection. $C_{\beta}(X;Y)$ and $K_{\beta}(X;Y)$
are discontinuous at $\beta=1,$ if there is common information between
the sources. Lemma \ref{lem:GCIproperties} implies Gács-Körner common
information, Wyner common information and exact common information
are extreme cases of $\beta$-approximate common information or $\beta$-exact
common information.
\end{rem}
\begin{IEEEproof}
To show (a), we only need to show for any variable $U$, there always
exists another variable $U'$ such that $|\mathcal{U}'|\leq|\mathcal{X}||\mathcal{Y}|+1,\rho_{m}(X;Y|U')=\rho_{m}(X;Y|U)$,
and $I(XY;U')=I(XY;U)$. Suppose $\rho_{m}(X;Y|U=u^{*})=\rho_{m}(X;Y|U)$.
According to Support Lemma \cite{El Gamal}, there exists a random
variable $U'$ with $\mathcal{U}'\subseteq\mathcal{U}$ and $|\mathcal{U}'|\leq|\mathcal{X}||\mathcal{Y}|+1$
such that
\begin{align}
P_{U'}(u^{*}) & =P_{U}(u^{*}),\label{eq:-58}\\
H(XY|U') & ={\displaystyle H(XY|U)},\\
P_{XY} & ={\displaystyle \sum_{u'}P_{U'}P_{XY|U'}}.\label{eq:-57}
\end{align}
\eqref{eq:-58} implies $\rho_{m}(X;Y|U')=\rho_{m}(X;Y|U)$. \eqref{eq:-57}
implies $H(XY)$ is also preserved, and hence $I(XY;U)=I(XY;U').$
This completes the proof of (a).

(b) \eqref{eq:-59} and \eqref{eq:-60} follow straightforwardly from
the definitions. According to the definitions and Lemma \ref{lem:relationship}
($\rho_{m}(X;Y|U)=0$ if and only if $X\rightarrow U\rightarrow Y$),
we can easily obtain \eqref{eq:-61} and \eqref{eq:-62}. Next we
prove \eqref{eq:-63}.

Consider
\begin{align}
C_{1^{-}}(X;Y) & ={\displaystyle \inf_{P_{U|X,Y}:\rho_{m}(X;Y|U)<1}I(XY;U)}.
\end{align}
Assume Gács-Körner common information is $f_{GK}(X,Y)$. Set $U=f_{GK}(X,Y)$,
then we have
\begin{align}
 & \rho_{m}(X;Y|U)<1,\\
 & I(XY;U)=H(f_{GK}(X,Y))=C_{GK}(X;Y).
\end{align}
Hence by definition,
\begin{equation}
C_{1^{-}}(X;Y)\leq C_{GK}(X;Y).\label{eq:-25}
\end{equation}

On the other hand, for any $U$ such that $\rho_{m}(X;Y|U)<1$, the
Gács-Körner common information is determined by $U$, i.e., $f_{GK}(X,Y)=g(U)$
for some function $g$. Therefore, we have
\begin{equation}
I(XY;U)=I(XY;U,f_{GK}(X,Y))\geq H(f_{GK}(X,Y))=C_{GK}(X;Y).
\end{equation}
Hence
\begin{equation}
C_{1^{-}}(X;Y)\geq C_{GK}(X;Y).\label{eq:-26}
\end{equation}

Combining \eqref{eq:-25} and \eqref{eq:-26} gives us
\begin{equation}
C_{1^{-}}(X;Y)=C_{GK}(X;Y).
\end{equation}

Similarly $K_{1^{-}}(X;Y)=C_{GK}(X;Y)$ can be proven as well.

(c) Suppose $P_{U|X,Y}$ achieves the infimum in \eqref{eq:-24}.
If $V$ satisfies both $XY\rightarrow U\rightarrow V$ and $XY\rightarrow V\rightarrow U$,
the we have $\rho_{m}(X;Y|U)=\rho_{m}(X;Y|UV)=\rho_{m}(X;Y|V)$.

If $V$ satisfies $XY\rightarrow U\rightarrow V$ but does not satisfy
$XY\rightarrow V\rightarrow U$, then $I(XY;U)=I(XY;UV)>I(XY;V)$.
Hence $\rho_{m}(X;Y|U)\leq\rho_{m}(X;Y|V)$, otherwise it contradicts
with that $P_{U|X,Y}$ achieves the infimum in \eqref{eq:-24}.
\end{IEEEproof}
Fig. \ref{fig:Illustration-of-the} illustrates the relationship among
joint entropy, mutual information, Gács-Körner common information,
Wyner common information, and generalized common information.

\begin{lem}
(Additivity and subadditivity). Assume $(X_{i},Y_{i})_{i=1}^{n}$
is a sequence of pairs of independent random variables, then we have
\begin{equation}
C_{\beta}(X^{n};Y^{n})={\displaystyle \sum_{i=1}^{n}C_{\beta}(X_{i};Y_{i})},\label{eq:-64}
\end{equation}
and
\begin{equation}
K_{\beta}(X_{i};Y_{i})\leq K_{\beta}(X^{n};Y^{n})\leq\sum_{i=1}^{n}K_{\beta}(X_{i};Y_{i}).\label{eq:-80}
\end{equation}
\end{lem}
\begin{IEEEproof}
For \eqref{eq:-64} it suffices to prove the $n=2$ case, i.e.,
\begin{equation}
C_{\beta}(X^{2};Y^{2})=C_{\beta}(X_{1};Y_{1})+C_{\beta}(X_{2};Y_{2}).\label{eq:-66}
\end{equation}

Observe for any $P_{U|X^{2}Y^{2}}$,
\begin{equation}
\rho_{m}(X^{2};Y^{2}|U)\geq\rho_{m}(X_{i};Y_{i}|U),i=1,2,
\end{equation}
and
\begin{align}
I(X^{2}Y^{2};U) & \geq I(X_{1}Y_{1};U)+I(X_{2}Y_{2};U|X_{1}Y_{1})\\
 & =I(X_{1}Y_{1};U)+I(X_{2}Y_{2};UX_{1}Y_{1})\\
 & \geq I(X_{1}Y_{1};U)+I(X_{2}Y_{2};U).
\end{align}
Hence we have
\begin{equation}
C_{\beta}(X^{2};Y^{2})\geq C_{\beta}(X_{1};Y_{1})+C_{\beta}(X_{2};Y_{2}).\label{eq:-65}
\end{equation}

Moreover,  if we choose $P_{U|X^{2}Y^{2}}=P_{U_{1}|X_{1}Y_{1}}^{*}P_{U_{2}|X_{2}Y_{2}}^{*}$
in $C_{\beta}(X^{2};Y^{2})$, where $P_{U_{i}|X_{i}Y_{i}}^{*},i=1,2,$
is the distribution achieving $C_{\beta}(X_{i};Y_{i})$, then we have
\begin{equation}
\rho_{m}(X^{2};Y^{2}|U)=\max_{i\in\left\{ 1,2\right\} }\rho_{m}(X_{i};Y_{i}|U_{i})\leq\beta,
\end{equation}
and
\begin{equation}
I(X^{2}Y^{2};U)=I(X_{1}Y_{1};U_{1})+I(X_{2}Y_{2};U_{2})=C_{\beta}(X_{1};Y_{1})+C_{\beta}(X_{2};Y_{2}).\label{eq:-65-1}
\end{equation}
Therefore,
\begin{equation}
C_{\beta}(X^{2};Y^{2})=\inf_{P_{U|X^{2}Y^{2}}:\rho_{m}(X^{2};Y^{2}|U)\leq\beta}I(X^{2}Y^{2};U)\leq C_{\beta}(X_{1};Y_{1})+C_{\beta}(X_{2};Y_{2}).\label{eq:-66-1}
\end{equation}

\eqref{eq:-65} and \eqref{eq:-66-1} implies \eqref{eq:-64} holds
for $n=2$.

Furthermore, the first inequality of \eqref{eq:-80} can be obtained
directly from the definition of $K_{\beta}$. The second inequality
of \eqref{eq:-80} can be obtained by restricting $P_{U|X^{n}Y^{n}}$
to the one with independent components (similar as the proof of \eqref{eq:-66-1}).
\end{IEEEproof}
For continuous sources, a lower bound on approximate common information
is given in the following theorem.
\begin{thm}
\label{thm:Lower-bound}(Lower bound on $C_{\beta}(X;Y)$). For any
continuous sources $\left(X,Y\right)$ with correlation coefficient
$\beta_{0},$ we have
\begin{equation}
C_{\beta}(X;Y){\displaystyle \geq h(XY)-\frac{1}{2}\log\left[(2\pi e(1-\beta_{0}))^{2}\frac{1+\beta}{1-\beta}\right]}
\end{equation}
for $0\leq\beta\leq\beta_{0}$, and $C_{\beta}(X;Y)=0$ for $\beta_{0}\leq\beta\leq1.$
\end{thm}
\begin{IEEEproof}
\begin{align}
I(XY;U) & =h(XY)-h(XY|U)\\
 & {\displaystyle \geq h(XY)-\mathbb{E}_{U}\frac{1}{2}\log\left[(2\pi e)^{2}\det(\Sigma_{XY|U})\right]}\\
 & {\displaystyle \geq h(XY)-\frac{1}{2}\log\left[(2\pi e)^{2}\det(\mathbb{E}_{U}\Sigma_{XY|U})\right]}\label{eq:-27}\\
 & =h(XY)-{\displaystyle \frac{1}{2}\log\left[(2\pi e)^{2}\left[\mathbb{E}\textrm{var}(X|U)\mathbb{E}\textrm{var}(Y|U)-(\mathbb{E}\textrm{cov}(X,Y|U))^{2}\right]\right]}\\
 & =h(XY)-{\displaystyle \frac{1}{2}\log\left[(2\pi e)^{2}\mathbb{E}\textrm{var}(X|U)\mathbb{E}\textrm{var}(Y|U)(1-\rho^{2}(X;Y|U))\right]}\\
 & {\displaystyle \geq h(XY)-\frac{1}{2}\log\left[(2\pi e)^{2}(\frac{1-\beta_{0}}{1-\rho(X,Y|U)})^{2}(1-\rho^{2}(X;Y|U))\right]}\label{eq:-28}\\
 & =h(XY)-{\displaystyle \frac{1}{2}\log\left[(2\pi e(1-\beta_{0}))^{2}\frac{1+\rho(X;Y|U)}{1-\rho(X;Y|U)}\right]}\\
 & {\displaystyle \geq h(XY)-\frac{1}{2}\log\left[(2\pi e(1-\beta_{0}))^{2}\frac{1+\beta}{1-\beta}\right]},\label{eq:-29}
\end{align}
where \eqref{eq:-27} follows from the function $\log(\det(\cdot))$
is concave on the set of symmetric positive definite square matrices
\cite[p.73]{Boyd}, \eqref{eq:-28} follows from Lemma \ref{lem:Conditioning-reduces-covariance},
and \eqref{eq:-29} follows from
\begin{equation}
\rho(X;Y|U)\leq\beta.
\end{equation}
\end{IEEEproof}
The equality holds in Theorem \ref{thm:Lower-bound} if $X,Y$ are
jointly Gaussian. The proof is given in Appendix \ref{sec:Proof-of-Theorem-Gaussian}.
\begin{thm}
\label{thm:Gaussian}(Gaussian sources). For jointly Gaussian sources
$X,Y$ with correlation coefficient $\beta_{0},$
\begin{equation}
C_{\beta}^{(G)}(X;Y)=\frac{1}{2}\log^{+}\left[\frac{1+\beta_{0}}{1-\beta_{0}}/\frac{1+\beta}{1-\beta}\right].
\end{equation}
\end{thm}
\begin{rem}
Specialized to the Wyner common information, $C_{W}^{(G)}(X;Y)=C_{0}^{(G)}(X;Y)={\displaystyle \frac{1}{2}\log^{+}\left[\frac{1+\beta_{0}}{1-\beta_{0}}\right]}$,
which was first given in \cite{Xu}.
\end{rem}
\begin{figure}
\begin{centering}
\includegraphics[width=0.6\textwidth]{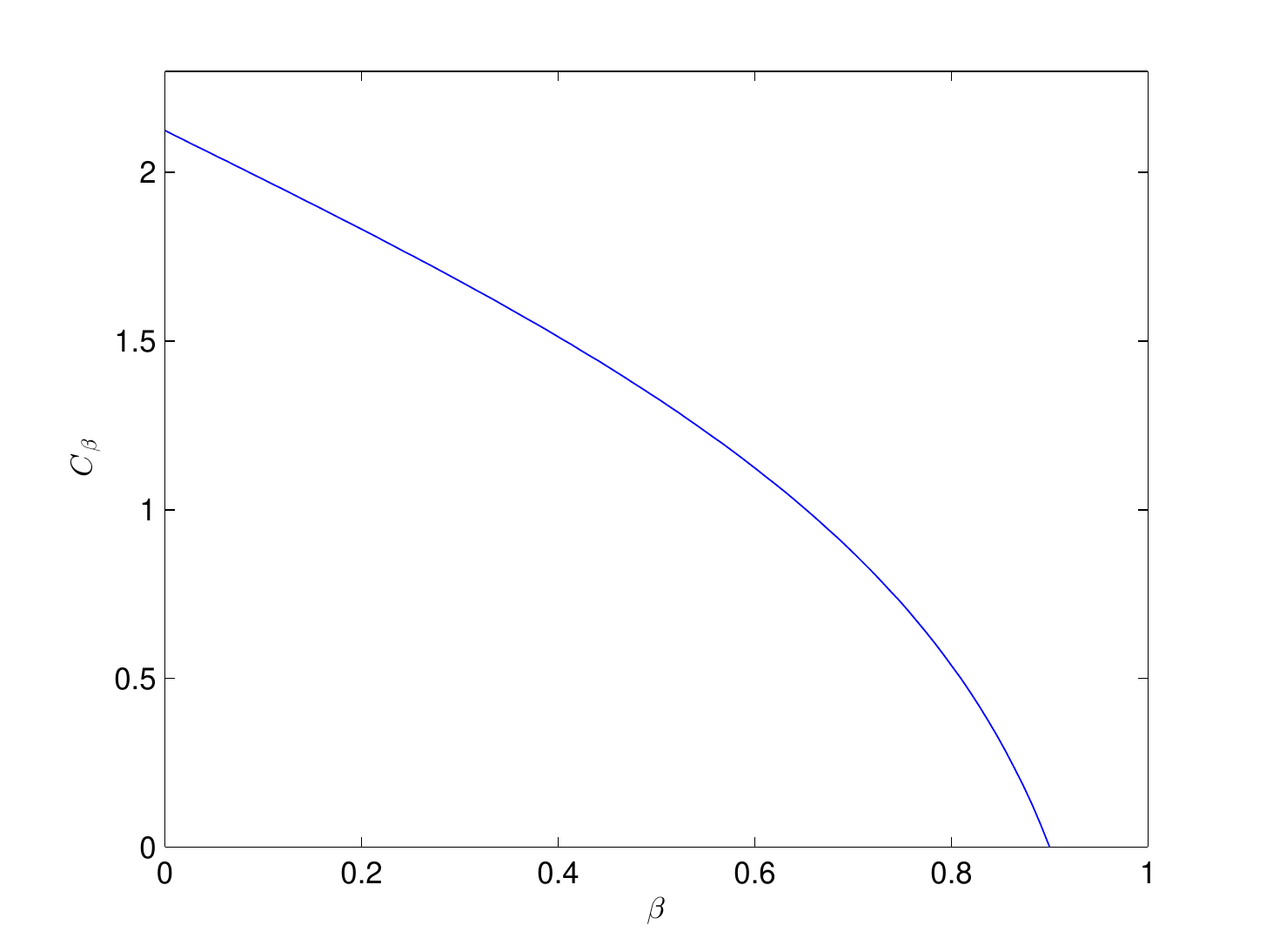}
\par\end{centering}
\caption{Information-correlation function for Gaussian sources in Theorem \ref{thm:Gaussian}
with $\beta_{0}=0.9.$}
\end{figure}

For the doubly symmetric binary source, an upper bound on common information
is given in the following theorem.
\begin{thm}
\label{thm:DSBS}(Doubly symmetric binary source (DSBS)). For doubly
symmetric binary source $\left(X,Y\right)$ with crossover probability
$p_{0},$  i.e., $P_{XY}=\left[\begin{array}{cc}
\frac{1}{2}\left(1-p_{0}\right) & \frac{1}{2}p_{0}\\
\frac{1}{2}p_{0} & \frac{1}{2}\left(1-p_{0}\right)
\end{array}\right],$ we have
\begin{equation}
C_{\beta}^{(B)}(X;Y)\leq1+H_{2}\left(p_{0}\right)-H_{4}\left(\frac{1}{2}\left(1-p_{0}+\sqrt{\frac{1-2p_{0}-\beta}{1-\beta}}\right),\frac{1}{2}\left(1-p_{0}-\sqrt{\frac{1-2p_{0}-\beta}{1-\beta}}\right),\frac{p_{0}}{2},\frac{p_{0}}{2}\right)\label{eq:-48}
\end{equation}
for $0\leq\beta<1-2p_{0}$, and $C_{\beta}^{(B)}(X;Y)=0$ for $\beta\geq1-2p_{0}$,
where  $H_{2}$ and $H_{4}$ denote the binary and quaternary entropy
functions, respectively, i.e.,
\begin{align}
H_{2}(p) & =-p\log p-(1-p)\log(1-p),\label{eq:binaryentropy}\\
H_{4}(a,b,c,d) & =-a\log a-b\log b-c\log c-d\log d.
\end{align}
\end{thm}
\begin{IEEEproof}
Assume $p$ is a value such that $2p\bar{p}=p_{0}$, $\bar{p}:=1-p$.
Then $(X,Y)$ can be expressed as
\begin{align}
X & =U\oplus V\oplus Z_{1},\\
Y & =U\oplus V\oplus Z_{2},
\end{align}
where $U\sim\textrm{Bern}(\frac{1}{2})$, $V\sim\textrm{Bern}(\alpha)$
with $0\leq\alpha\leq1$, $Z_{1}\sim\textrm{Bern}(p)$, and $Z_{2}\sim\textrm{Bern}(p)$
are independent. Hence we have $P_{V\oplus Z_{1},V\oplus Z_{2}}=\left[\begin{array}{cc}
a & p\bar{p}\\
p\bar{p} & b
\end{array}\right]$ with $a=\alpha p^{2}+\bar{\alpha}\bar{p}^{2},b=\alpha\bar{p}^{2}+\bar{\alpha}p^{2}$,
and

\begin{equation}
\rho_{m}(X,Y|U)=\rho_{m}(V\oplus Z_{1},V\oplus Z_{2}).
\end{equation}
By using the formula
\begin{equation}
\rho_{m}^{2}(X,Y)\leq\left[\sum_{x,y}\frac{P^{2}(x,y)}{P(x)P(y)}\right]-1
\end{equation}
for $(X,Y)$ with at least one of them being binary-valued,  we have
\begin{equation}
\rho_{m}(X,Y)=1-2p_{0}.
\end{equation}
Hence $C_{\beta}^{(B)}(X;Y)=0$ for $\beta\geq1-2p_{0}$. Next we
consider the case
\begin{equation}
\beta\leq1-2p_{0}.
\end{equation}
To guarantee $\rho_{m}(V\oplus Z_{1},V\oplus Z_{2})\leq\beta$, we
choose
\begin{align}
a & =\frac{1}{2}\left(1-p_{0}+\sqrt{\frac{1-2p_{0}-\beta}{1-\beta}}\right)\\
b & =\frac{1}{2}\left(1-p_{0}-\sqrt{\frac{1-2p_{0}-\beta}{1-\beta}}\right).
\end{align}
This leads to the inequality \eqref{eq:-45}. This completes the
proof.

\end{IEEEproof}

\subsection{\textit{Relationship to Rate-Distortion Function}}

The approximate information-correlation function can be rewritten
as
\begin{equation}
C_{\beta}(X;Y)={\displaystyle \inf_{P_{U|X,Y}:d(P_{UXY})\leq\beta}I(XY;U)}.
\end{equation}
where $d(P_{UXY}):=\rho_{m}(X;Y|U)$. This expression has a form similar
to rate-distortion function, if we consider maximal correlation as
a special ``distortion measure''. But it is worth nothing that maximal
correlation is taken on the distribution of $X,Y$, instead of on
them itself.

Information-correlation function is also related to the rate-privacy
function \cite{Asoodeh}
\begin{equation}
g_{\beta}(X;Y):={\displaystyle \sup_{P_{U|Y}:\rho_{m}(X;U)\leq\beta}I(Y;U)},
\end{equation}
in which $U$ can be thought of as the extracted information from
$Y$ under privacy constraint $\rho_{m}(X;U)\leq\beta$. But there
are three differences between $g_{\beta}(X;Y)$ and $C_{\beta}(X;Y)$.
1) The privacy constraint in $g_{\beta}(X;Y)$ is a constraint on
unconditional maximal correlation, and moreover, this unconditional
maximal correlation is that between the remote source $X$ and extracted
information $U$, instead of between the sources. Hence $g_{\beta}(X;Y)$
is not symmetric respect to $X,Y$. 2) In $g_{\beta}(X;Y)$, $U$
is extracted from $Y$ instead of both $X,Y$, hence $X\rightarrow Y\rightarrow U$
is restricted in $g_{\beta}(X;Y)$. 3) The optimization in $C_{\beta}(X;Y)$
is infimum, while in $g_{\beta}(X;Y)$ is supremum.

\section{Private Sources Synthesis}

In order to provide an  operational interpretation for information-correlation
functions $C_{\beta}(X;Y)$ and $K_{\beta}(X;Y)$, in this section,
we consider \emph{private sources synthesis problem}. We show that
the information-correlation functions correspond to the minimum achievable
rates for the centralized setting version of this problem.

\subsection{\textit{Problem Setup}}

\begin{figure}
\begin{centering}
\includegraphics[width=0.3\textwidth]{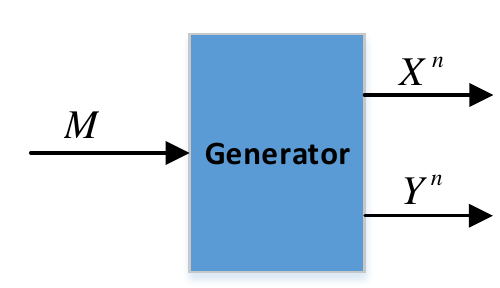}\qquad{}\includegraphics[width=0.3\textwidth]{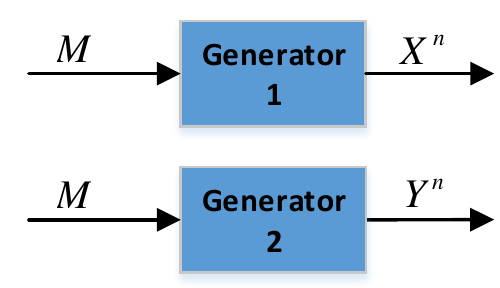}
\par\end{centering}
\caption{\label{fig:Private-source-synthesis}Private source synthesis problem:
(left) centralized setting; (right) distributed setting. In this problem
we assume 1) rate constraint  $\limsup_{n\rightarrow\infty}{\displaystyle \frac{1}{n}H(M)\leq R}$;
2) privacy constraint $\rho_{m}(X^{n};Y^{n}|M)\leq\beta$; 3) source
distribution constraint ${\displaystyle \lim_{n\rightarrow\infty}\Vert P_{X^{n}Y^{n}}-Q_{X^{n}Y^{n}}\Vert_{TV}=0}$
in approximate synthesis sense, or $P_{X^{n}Y^{n}}=Q_{X^{n}Y^{n}}$
in exact synthesis sense. For distributed setting, the $M$ in the
constraints is replaced with $M_{1}M_{2}$.}
\end{figure}

Consider private sources synthesis problem shown in Fig. \ref{fig:Private-source-synthesis},
where a simulator generates two source sequences $X^{n}$ and $Y^{n}$
from a common random variable $M$. $X^{n}$ and $Y^{n}$ are restricted
to follow i.i.d. according to a target distribution ${\displaystyle \prod P_{XY}.}$
\begin{defn}
A generator is defined by a pmf $P_{M}$ and a stochastic mapping
 $P_{X^{n}Y^{n}|M}:\mathcal{M}\mapsto\mathcal{X}^{n}\times\mathcal{Y}^{n}$.
\end{defn}
Furthermore, Shannon's zero-error source coding theorem states that,
it is possible to compress a message $M$ (using a variable length
coding) at rate ${\displaystyle R}$ for sufficiently large $n$ if
$R>\frac{1}{n}H(M)$; and conversely, it is possible only if $R\geq\frac{1}{n}H(M)$.
Hence we define the achievability of tuple $(R,\beta)$ as follows.
\begin{defn}
The tuple $(R,\beta)$ is approximately or exactly achievable if there
exists a sequence of generators such that\\
1) rate constraint:
\begin{equation}
\limsup_{n\rightarrow\infty}\frac{1}{n}H(M)\leq R;\label{eq:-30-1}
\end{equation}
2) privacy constraint:
\begin{equation}
\rho_{m}(X^{n};Y^{n}|M)\leq\beta,\forall n;\label{eq:-30}
\end{equation}
3) approximate sources distribution constraint:
\end{defn}
\begin{equation}
{\displaystyle \lim_{n\rightarrow\infty}\Vert P_{X^{n}Y^{n}}-\prod P_{XY}\Vert_{TV}=0},\label{eq:-31}
\end{equation}
or exact sources distribution constraint:
\begin{equation}
P_{X^{n}Y^{n}}={\displaystyle \prod P_{XY}},\forall n.\label{eq:-32}
\end{equation}

\begin{defn}
The rate-correlation function for approximate private sources synthesis
is defined by $R_{PSS}(\beta):=\inf\left\{ R:(R,\beta)\textrm{ is approximately achievable}\right\} $.
Similarly, the rate-correlation function for exact private sources
synthesis is defined by $R_{PSS}^{(E)}(\beta):=\inf\left\{ R:(R,\beta)\textrm{ is exactly achievable}\right\} $.
\end{defn}
Furthermore, we also consider distributed setting, which is shown
in Fig. \ref{fig:Private-source-synthesis} (b). For this case, the
source synthesis problem is named \emph{distributed private sources
synthesis}.
\begin{defn}
A distributed generator is defined by a pmf $P_{M}$ and two stochastic
mappings: $P_{X^{n}|M}:\mathcal{M}\mapsto\mathcal{X}^{n}$ and $P_{Y^{n}|M}:\mathcal{M}\mapsto\mathcal{Y}^{n}$.
\end{defn}
\begin{defn}
The tuple $(R,\beta)$ is approximately or exactly achievable for
distributed setting if there exists a sequence of distributed generators
such that\\
1) rate constraint: \eqref{eq:-30-1};\\
2) privacy constraint: \eqref{eq:-30};\\
3) approximate source distribution constraint: \eqref{eq:-31}, or
exact source distribution constraint: \eqref{eq:-32}.
\end{defn}
\begin{defn}
The rate-correlation function for distributed approximate or exact
private sources synthesis is defined by $R_{DPSS}(\beta):=\inf\left\{ R:(R,\beta)\textrm{ is approximately achievable}\right\} $
and $R_{DPSS}^{(E)}(\beta):=\inf\left\{ R:(R,\beta)\textrm{ is exactly achievable}\right\} $,
respectively.
\end{defn}
For distributed setting, privacy constraint
\begin{equation}
\rho_{m}(X^{n};Y^{n}|M)=0
\end{equation}
is satisfied immediately. Therefore,
\begin{align}
R_{DPSS}(\beta) & =R_{DPSS}(0),\\
R_{DPSS}^{(E)}(\beta) & =R_{DPSS}^{(E)}(0).
\end{align}

We assume the synthesized sources have finite alphabets.

\subsection{\textit{Main Result}}

\subsubsection{\textit{Centralized Setting}}

For approximate private sources synthesis, we have the following theorems.
The proof of Theorem \ref{thm:app-cent-PSS} is given in Appendix
\ref{sec:Proof-of-Theorem-PSS}.
\begin{thm}
\label{thm:app-cent-PSS}For approximate private sources synthesis,
\begin{equation}
R_{PSS}(\beta)=C_{\beta}(X;Y).
\end{equation}
\end{thm}
\begin{rem}
From the proof we can see that using fixed-length coding is sufficient
to achieve the rate-correlation function $R_{PSS}(\beta)$.
\end{rem}
\begin{thm}
For exact private sources synthesis,
\begin{equation}
R_{PSS}^{(E)}(\beta)=K_{\beta}(X;Y).
\end{equation}
\end{thm}
\begin{IEEEproof}
\emph{Achievability:} Suppose $R>K_{\beta}(X;Y)$. We will show that
the rate $R$ is achievable.

\textbf{Input Process Generator:} Generate input source $M$ according
to pmf $P_{U_{n}}.$

\textbf{Source Generator:}\emph{ }Upon $m$, the generator generate
sources $(X^{n},Y^{n})$ according to ${\displaystyle P_{X^{n}Y^{n}|U_{n}}(x^{n},y^{n}|m)}$.

For such generator, the induced overall distribution is
\begin{equation}
P_{X^{n}Y^{n}M}(x^{n},y^{n},m):=P_{X^{n}Y^{n}U_{n}}(x^{n},y^{n},m).
\end{equation}
This means
\begin{equation}
\rho_{m}(X^{n};Y^{n}|M)\leq\beta,
\end{equation}
since
\begin{equation}
\rho_{m}(X^{n};Y^{n}|U_{n})\leq\beta.
\end{equation}

Since ${\displaystyle K_{\beta}(X;Y)=\lim_{n\rightarrow\infty}\frac{1}{n}H(U_{n})}$
for some $U_{n}$, $R{\displaystyle \geq\frac{1}{n}(H(U_{n})+1)}$
for $n$ large enough. By the achievability part of Shannon's zero-error
source coding theorem, it is possible to exactly generate $(X^{n},Y^{n})$
at rate at most ${\displaystyle \frac{1}{n}(H(U_{n})+1)}$. Hence
rate $R$ is achievable and thus $R_{PSS}^{(E)}(\beta)\leq K_{\beta}(X;Y)$.

\emph{Converse: }Now suppose a rate $R$ is achievable. Then there
exists an $(n,R)$-generator that exactly generates $(X^{n},Y^{n})$
such that
\begin{equation}
\rho_{m}(X^{n};Y^{n}|M)\leq\beta.
\end{equation}
By the converse for Shannon's zero-error source coding theorem,
\begin{equation}
{\displaystyle \lim_{n\rightarrow\infty}\frac{1}{n}H(M)\leq R}.
\end{equation}
Therefore,
\begin{equation}
R{\displaystyle \geq\lim_{n\rightarrow\infty}\frac{1}{n}H(M)\geq\lim_{n\rightarrow\infty}\inf_{P_{U_{n}|X^{n}Y^{n}}:\rho_{m}(X^{n},Y^{n}|U_{n})\leq\beta}\frac{1}{n}H(U_{n})=K_{\beta}(X;Y)}.
\end{equation}
That is
\begin{equation}
R_{PSS}^{(E)}(\beta)\geq K_{\beta}(X;Y).
\end{equation}
\end{IEEEproof}

\subsubsection{\textit{Distributed Setting}}

For distributed private sources synthesis, we have similar results.
\begin{thm}
For distributed approximate private sources synthesis,
\begin{equation}
R_{DPSS}(\beta)=C_{0}(X;Y).
\end{equation}
\end{thm}
\begin{rem}
From the proof we can see that similar to centralized case, using
fixed-length coding is also sufficient to achieve the rate-correlation
function $R_{DPSS}(\beta)$ for distributed case.
\end{rem}
\begin{IEEEproof}
The theorem was essentially same to Wyner's result \cite{Wyner}.
In the following, we prove this theorem by following similar steps
to the proof of the centralized case.

\emph{Achievability: }Consider the generator used for the centralized
case (see Appendix \ref{subsec:PSSAchievability}). Similar to the
centralized case, we can prove if $R>C_{0}(X;Y)$,
\begin{equation}
{\displaystyle \lim_{n\rightarrow\infty}\mathbb{E}_{\mathcal{C}}\Vert P_{X^{n}Y^{n}}-Q_{X^{n}Y^{n}}\Vert_{TV}=0}.
\end{equation}
Owing to the distributed setting, Markov chain $X^{n}\rightarrow M\rightarrow Y^{n}$
holds. By Lemma \ref{lem:relationship}, we have
\begin{equation}
\rho_{m}(X^{n};Y^{n}|M)=0.
\end{equation}
Hence
\begin{equation}
R_{DPSS}(\beta)\leq C_{0}(X;Y).
\end{equation}

\emph{Converse: }By slightly modified the proof of centralized case
and combining with Markov chain $X^{n}\rightarrow M\rightarrow Y^{n}$,
we can show that
\begin{equation}
R_{DPSS}(\beta)\geq C_{0}(X;Y).
\end{equation}
\end{IEEEproof}
\begin{thm}
For distributed exact private sources synthesis,
\begin{equation}
R_{DPSS}^{(E)}(\beta)=K_{0}(X;Y).
\end{equation}
\end{thm}
\begin{IEEEproof}
\emph{Achievability:} Suppose $R>K_{0}(X;Y)$. We will show that the
rate $R$ is achievable.

\textbf{Input Process Generator:} Generate input source $M$ according
to $P_{U_{n}}.$

\textbf{Source Generator: }Upon $m$, the generator 1 generates source
$X^{n}$ according to ${\displaystyle P_{X^{n}|U_{n}}(x^{n}|m)}$,
and the generator 2 generates source $Y^{n}$ according to ${\displaystyle P_{Y^{n}|U_{n}}(y^{n}|m)}$.

Similar to the centralized case, since $\rho_{m}(X^{n};Y^{n}|U_{n})=0$,
i.e., $X^{n}\rightarrow U_{n}\rightarrow Y^{n}$, the induced overall
distribution is
\begin{equation}
P_{X^{n}Y^{n}M}(x^{n},{\displaystyle y^{n},m):=P_{U_{n}}\left(m\right){\displaystyle P_{X^{n}|U_{n}}(x^{n}|m)}{\displaystyle P_{Y^{n}|U_{n}}(y^{n}|m)}=P_{X^{n}Y^{n}U_{n}}(x^{n},y^{n},m)}.
\end{equation}
This means
\begin{equation}
P_{X^{n}Y^{n}}(x^{n},{\displaystyle y^{n})=\prod_{i=1}^{n}P_{XY}(x_{i},y_{i})}.
\end{equation}
and
\begin{equation}
\rho_{m}(X^{n};Y^{n}|M)=0\leq\beta.
\end{equation}
Hence the rate $R$ is achievable, which further implies
\begin{equation}
R_{DPSS}^{(E)}(\beta)\leq K_{0}(X;Y).
\end{equation}

\emph{Converse:} Suppose a rate $R$ is achievable. Then there exists
an $(n,R)$-generator that exactly generates $(X^{n},Y^{n})$ such
that
\begin{equation}
\rho_{m}(X^{n};Y^{n}|M)\leq\beta.
\end{equation}
Owing to the distributed setting, Markov chain $X^{n}\rightarrow M\rightarrow Y^{n}$
holds naturally. By Lemma \ref{lem:relationship}, we have
\begin{equation}
\rho_{m}(X^{n};Y^{n}|M)=0.
\end{equation}
Furthermore, by the converse for Shannon's zero-error source coding
theorem,
\begin{equation}
{\displaystyle \lim_{n\rightarrow\infty}\frac{1}{n}H(M)\leq R}.
\end{equation}
Therefore,
\begin{equation}
R{\displaystyle \geq\lim_{n\rightarrow\infty}\frac{1}{n}H(M)\geq\lim_{n\rightarrow\infty}\inf_{P_{U_{n}|X^{n}Y^{n}}:\rho_{m}(X^{n},Y^{n}|U_{n})\leq\beta}\frac{1}{n}H(U_{n})=K_{0}(X;Y)}.
\end{equation}
That is
\begin{equation}
R_{DPSS}^{(E)}(\beta)\geq K_{0}(X;Y).
\end{equation}
\end{IEEEproof}

\section{\label{sec:Common-Information-Extraction}Common Information Extraction}

In this section, we study another problem, \emph{common information
extraction problem}, which provides another operational interpretation
for information-correlation functions $C_{\beta}(X;Y)$ and $K_{\beta}(X;Y)$.
Similar to private sources synthesis problem, the information-correlation
functions are proven to be the minimum achievable rates for the centralized
setting version of this problem as well.

\subsection{\textit{Problem Setup}}

\begin{figure}
\begin{centering}
\includegraphics[width=0.3\textwidth]{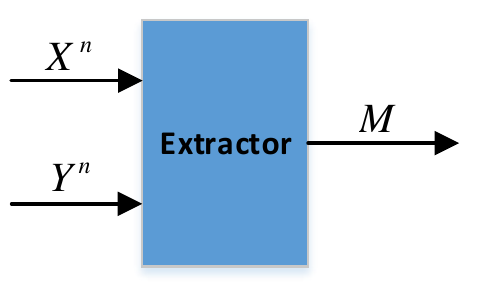}\qquad{}\includegraphics[width=0.3\textwidth]{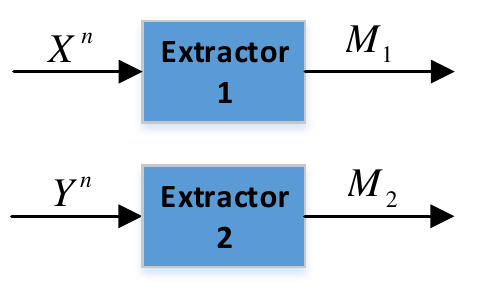}
\par\end{centering}
\caption{\label{fig:Common-information-extraction}Common information extraction
problem: (left) centralized setting; (right) distributed setting.
In this problem we assume 1) rate constraint  ${\displaystyle \limsup_{n\rightarrow\infty}\frac{1}{n}H(M)\leq R}$;
2) weak privacy constraint: for any $\epsilon>0$, ${\displaystyle {\displaystyle \widetilde{\rho}_{m}^{\epsilon}(X^{n};Y^{n}|M)\leq\beta},\forall n,}$
or strong privacy constraint: $\rho_{m}(X^{n};Y^{n}|M)\leq\beta,\forall n.$
For distributed setting, the variable $M$ in the constraints is replaced
with $M_{1}M_{2}$.}
\end{figure}

As a counterpart of private sources synthesis problem, we consider
common information extraction problem shown in Fig. \ref{fig:Common-information-extraction},
where an extractor extracts common random variable $M$ from two source
sequences $X^{n}$ and $Y^{n}.$ $X^{n}$ and $Y^{n}$ are i.i.d.
according to $P_{XY}.$

\begin{defn}
An extractor is defined by a stochastic mapping: $P_{M|X^{n}Y^{n}}:\mathcal{X}^{n}\times\mathcal{Y}^{n}\mapsto\mathcal{M}$.
\end{defn}
The extractor should extract an enough mount of common information
to satisfy the privacy constraint measured by conditional maximal
correlation.
\begin{defn}
The tuple $(R,\beta)$ is weakly or strongly achievable if there exists
a sequence of extractors such that\\
1) rate constraint:
\begin{equation}
\limsup_{n\rightarrow\infty}\frac{1}{n}H(M)\leq R;\label{eq:-30-1-1}
\end{equation}
2a) weak privacy constraint:   for any $\epsilon>0$, it holds that
\begin{equation}
{\displaystyle \widetilde{\rho}_{m}^{\epsilon}(X^{n};Y^{n}|M)\leq\beta},\forall n,
\end{equation}
where $\widetilde{\rho}_{m}^{\epsilon}(X^{n};Y^{n}|M)$ denotes $\epsilon$-smooth
conditional maximal correlation; see \eqref{eq:-81}; \\
2b) or strong privacy constraint:
\begin{equation}
\rho_{m}(X^{n};Y^{n}|M)\leq\beta,\forall n.
\end{equation}
\end{defn}
Common information corresponds to the smallest information rate that
makes the privacy constraint satisfied, hence the common information
indeed represents a kind of ``core\textquotedblright{} information.

Now we define the rate-correlation functions as follows.
\begin{defn}
The rate-correlation functions for weakly and strongly common information
extraction problems are defined by $R_{CIE}(\beta):=\inf\left\{ R:(R,\beta)\textrm{ is weakly achievable}\right\} $
and $R_{CIE}^{(E)}(\beta):=\inf\left\{ R:(R,\beta)\textrm{ is strongly achievable}\right\} $,
respectively.
\end{defn}
Furthermore, we also consider distributed common information extraction.
\begin{defn}
A distributed extractor is defined by two stochastic mappings $P_{M_{1}|X^{n}}:\mathcal{X}^{n}\mapsto\mathcal{M}_{1}$
and $P_{M_{2}|Y^{n}}:\mathcal{Y}^{n}\mapsto\mathcal{M}_{2}$.
\end{defn}
\begin{defn}
The tuple $(R,\beta)$ is achievable for distributed setting if there
exists a sequence of distributed extractors such that\\
1) rate constraint: \eqref{eq:-30-1-1};\\
2a) weak privacy constraint: for any $\epsilon>0$, it holds that
\begin{equation}
{\displaystyle \widetilde{\rho}_{m}^{\epsilon}(X^{n};Y^{n}|M_{1},M_{2})\leq\beta},\forall n,
\end{equation}
2b) or strong privacy constraint:
\begin{equation}
\rho_{m}(X^{n};Y^{n}|M_{1},M_{2})\leq\beta,\forall n.
\end{equation}
\end{defn}
\begin{defn}
The rate-correlation functions for distributed weakly and strongly
common information extraction problems are defined by $R_{DCIE}(\beta):=\inf\left\{ R:(R,\beta)\textrm{ is weakly achievable}\right\} $
and $R_{DCIE}^{(E)}(\beta):=\inf\{R:(R,\beta)\textrm{ is strongly achievable}\}$,
respectively.
\end{defn}
We also assume the sources have finite alphabets.

\subsection{\textit{Main Result}}

\subsubsection{\textit{Centralized Setting}}

For weakly common information extraction, we have the following theorems.
The proof of Theorem \ref{thm:app-cent-CIE} is given in Appendix
\ref{sec:Proof-of-Theorem-CIE}.
\begin{thm}
\label{thm:app-cent-CIE}For weakly common information extraction,
\begin{equation}
R_{CIE}(\beta)=C_{\beta}(X;Y).
\end{equation}
\end{thm}
\begin{rem}
From the proof we can see that using fixed-length coding is sufficient
to achieve the rate-correlation function $R_{CIE}(\beta)$.
\end{rem}
\begin{thm}
For strongly common information extraction,
\begin{equation}
R_{CIE}^{(E)}(\beta)=K_{\beta}(X;Y).
\end{equation}
\end{thm}
\begin{IEEEproof}
\emph{Achievability:} Suppose $R>K_{\beta}(X;Y)$. We will show that
the rate $R$ is achievable.

\textbf{Extractor:} Upon $(x^{n},y^{n})$, the extractor generates
$m$ according to $P_{U_{n}|X^{n}Y^{n}}(m|x^{n},y^{n})$.

For such extractor, the induced overall distribution is
\begin{equation}
P_{X^{n}Y^{n}M}(x^{n},y^{n},m)=P_{X^{n}Y^{n}U_{n}}(x^{n},y^{n},m).
\end{equation}
Hence
\begin{equation}
\rho_{m}(X^{n};Y^{n}|M)=\rho_{m}(X^{n};Y^{n}|U_{n})\leq\beta.
\end{equation}
Since ${\displaystyle K_{\beta}(X;Y)=\lim_{n\rightarrow\infty}\frac{1}{n}H(U_{n}),}$
$R{\displaystyle \geq\frac{1}{n}(H(U_{n})+1)}$ for $n$ large enough.
By the achievability part of Shannon's zero-error source coding theorem,
it is possible to exactly generate $(X^{n},Y^{n})$ at rate at most
${\displaystyle \frac{1}{n}(H(U_{n})+1)}$. Hence rate $R$ is achievable
and thus $R_{CIE}^{(E)}(\beta)\leq K_{\beta}(X;Y)$.

\emph{Converse:} Now suppose a rate $R$ is achievable. Then there
exists a sequence of extractors that generate $M$ such that
\begin{equation}
\rho_{m}(X^{n};Y^{n}|M)\leq\beta,\forall n.
\end{equation}
By the converse for Shannon's zero-error source coding theorem,
\begin{equation}
{\displaystyle \lim_{n\rightarrow\infty}\frac{1}{n}H(M)\leq R}.
\end{equation}
Therefore,
\begin{equation}
R{\displaystyle \geq\lim_{n\rightarrow\infty}\frac{1}{n}H(M)\geq\lim_{n\rightarrow\infty}\inf_{P_{U_{n}|X^{n}Y^{n}}:\rho_{m}(X^{n},Y^{n}|U_{n})\leq\beta}\frac{1}{n}H(U_{n})=K_{\beta}(X;Y)}.
\end{equation}
That is
\begin{equation}
R_{CIE}^{(E)}(\beta)\geq K_{\beta}(X;Y).
\end{equation}
\end{IEEEproof}

\subsubsection{\textit{Distributed Setting}}

For distributed common information extraction, we have similar results.
The following theorems hold for weakly and strongly common information
extraction, respectively. The proof of Theorem \ref{thm:app-dist-CIE}
is given in Appendix \ref{sec:Proof-of-Theorem-CIE-dist}.
\begin{thm}
\label{thm:app-dist-CIE}For distributed weakly common information
extraction,
\begin{equation}
C_{\beta}^{(D,LB)}(X;Y)\leq R_{DCIE}(\beta)=C_{\beta}^{(D)}(X;Y)\leq C_{\beta}^{(D,UB)}(X;Y),
\end{equation}
where
\begin{align}
C_{\beta}^{(D,UB)}(X;Y) & :=\inf_{P_{U|X}P_{V|Y}:\rho_{m}(X,Y|UV)\leq\beta}{\displaystyle I(XY;UV)},\\
C_{\beta}^{(D)}(X;Y) & :=\lim_{n\rightarrow\infty}\inf_{P_{U|X^{n}}P_{V|Y^{n}}:\rho_{m}(X^{n};Y^{n}|UV)\leq\beta}\frac{1}{n}{\displaystyle I(X^{n}Y^{n};UV)},\\
C_{\beta}^{(D,LB)}(X;Y) & :=\inf_{\substack{P_{T}P_{UV|XYT}:UT\rightarrow X\rightarrow Y,X\rightarrow Y\rightarrow VT,\\
\rho_{m}(UX;VY|T)\leq\rho_{m}(X;Y),\\
\rho_{m}(X,Y|UVT)\leq\beta
}
}{\displaystyle I(XY;UV|T)}.
\end{align}
\end{thm}
\begin{rem}
From the proof we can see that similar to centralized case, using
fixed-length coding is also sufficient to achieve the rate-correlation
function $R_{DCIE}(\beta)$ for distributed case.
\end{rem}
\begin{thm}
For distributed strongly common information extraction,
\begin{equation}
R_{DCIE}^{(E)}(\beta)=K_{\beta}^{(D)}(X;Y),
\end{equation}
where
\begin{equation}
K_{\beta}^{(D)}(X^{n};Y^{n}):=\lim_{n\rightarrow\infty}{\displaystyle \inf_{P_{U_{n}|X^{n}}P_{V_{n}|Y^{n}}:\rho_{m}(X^{n},Y^{n}|U_{n}V_{n})\leq\beta}\frac{1}{n}H(U_{n}V_{n})}.
\end{equation}
\end{thm}
\begin{IEEEproof}
\emph{Achievability:} Suppose $R>K_{\beta}^{(D)}(X;Y)$. We will show
that the rate $R$ is achievable.

\textbf{Extractor: }Upon $(x^{n},Y^{n})$, the extractor 1 generates
$m_{1}$ according to $P_{U_{n}|X^{n}}(m_{1}|x^{n})$, and extractor
2 generates $m_{2}$ according to $P_{V_{n}|Y^{n}}(m_{2}|y^{n})$.

For such extractor, the induced overall distribution is
\begin{equation}
P_{X^{n}Y^{n}M_{1}M_{2}}(x^{n},Y^{n},m_{1},m_{2})=P_{X^{n}Y^{n}U_{n}V_{n}}(x^{n},Y^{n},m_{1},m_{2}).
\end{equation}
Hence
\begin{equation}
\rho_{m}(X^{n};Y^{n}|M_{1}M_{2})=\rho_{m}(X^{n};Y^{n}|U_{n}V_{n})\leq\beta.
\end{equation}
Since ${\displaystyle \overline{G}_{\beta}(X;Y)=\lim_{n\rightarrow\infty}\frac{1}{n}H(U_{n}V_{n}),}$
$R{\displaystyle \geq\frac{1}{n}(H(U_{n}V_{n})+1)}$ for $n$ large
enough. By the achievability part of Shannon's zero-error source coding
theorem, it is possible to exactly generate $(X^{n},Y^{n})$ at rate
at most ${\displaystyle \frac{1}{n}(H(U_{n}V_{n})+1)}$. Hence rate
$R$ is achievable and thus $R_{DCIE}^{(E)}(\beta)\leq K_{\beta}^{(D)}(X;Y)$.

\emph{Converse: }Now suppose a rate $R$ is achievable. Then there
exists a sequence of extractors that generate $(M_{1},M_{2})$ such
that
\begin{equation}
\rho_{m}(X^{n};Y^{n}|M_{1}M_{2})\leq\beta,\forall n.
\end{equation}
By the converse for Shannon's zero-error source coding theorem,
\begin{equation}
{\displaystyle \lim_{n\rightarrow\infty}\frac{1}{n}H(M_{1}M_{2})\leq R}.
\end{equation}
Therefore,
\begin{equation}
R{\displaystyle \geq\lim_{n\rightarrow\infty}\frac{1}{n}H(M_{1}M_{2})\geq\lim_{n\rightarrow\infty}\inf_{p_{U_{n}|X^{n}}p_{V_{n}|Y^{n}}:\rho_{m}(X^{n},Y^{n}|U_{n}V_{n})\leq\beta}\frac{1}{n}H(U_{n}V_{n})=K_{\beta}^{(D)}(X;Y)}.
\end{equation}
That is
\begin{equation}
R_{DCIE}^{(E)}(\beta)\geq K_{\beta}^{(D)}(X;Y).
\end{equation}
\end{IEEEproof}

\section{Concluding Remarks}

In this paper, we unify and generalize Gács-Körner and Wyner common
informations, and define a generalized version of common information,
(approximate) information-correlation function, by exploiting maximal
correlation as a commonness or privacy measure. The Gács-Körner common
information and Wyner common information are two special and extreme
cases of our generalized definition. Furthermore, similarly exact
information-correlation function has been defined as well, which is
a generalization of Gács-Körner common information and Kumar-Li-Gamal
common information. We study the problems of common information extraction
and private sources synthesis, and show that these two information-correlation
functions are equal to the optimal rates under given correlation constraints
in the centralized cases of these problems.

Our results have a sequence of applications:
\begin{itemize}
\item Dependency measure: The generalized common informations defined by
us provide a fresh look at dependency. The more common information
the sources share, the more dependent they are. To normalize the (approximate)
information-correlation function, we can define
\begin{equation}
{\displaystyle \Gamma_{\beta}(X;Y)=\frac{C_{\beta}(X;Y)}{H(X,Y)}},
\end{equation}
or
\begin{equation}
\Gamma_{\beta}(X;Y)=1-2^{-2C_{\beta}(X;Y)}.
\end{equation}
Furthermore, we define \emph{correlation-information function} as
the inverse function of information-correlation function, i.e.,
\begin{equation}
\beta_{C}(X;Y)={\displaystyle \inf_{P_{U|XY}:I(XY;U)\leq C}}\rho_{m}(X;Y|U),\label{eq:-24-3}
\end{equation}
which represents the source dependency after extracting $C$-rate
common information from $X,Y$. Obviously $\beta_{C}(X;Y)=\rho_{m}(X;Y)$
when $C=0$. Dependency measure can be further applied to feature
extraction and image classification. Furthermore, conditional maximal
correlation can be also applied to measure the dependency of distributed
sources, which has been exploited to derive some converse results
of distributed communication; see our another work \cite{Yu}.
\item Game theory and correlation based secrecy: The common information
extraction can be equivalently transformed into a zero-sum game problem.
Consider two adversarial parties. One is Player A, and another one
is Players B and C. Players A and B share a source $X$, and Players
A and C share another source $Y$. Sources $X,Y$ are correlated and
memoryless. Players B and C cooperate to maximize the conditional
correlation $\rho(f(X^{n},M);g(Y^{n},M)|M)$ (or $\rho_{Q}(f(X^{n},M);g(Y^{n},M)|M)$
for some distribution $Q_{X^{n}Y^{n}M}$) over all functions $f,g$,
where $M$ is a message received from Player A through a rate-limited
channel, and $f(X^{n},M)$ and $g(Y^{n},M)$ are the outputs of Players
B and C respectively. Player A generates $M$ from $X^{n},Y^{n}$
and wants to minimize the optimal correlation induced by Players
B and C (assume Player A does not know the distribution $Q$ Players
A and B choose). Then our result on common information extraction
can directly apply to this case, and it implies the exact (or approximate)
information-correlation function is equal to the minimum rate needed
for Player A to force B and C's optimal strategy satisfying  $\sup_{f,g}\rho(f(X^{n},M);g(Y^{n},M)|M)\leq\beta$
(or $\inf_{\left\Vert Q_{X^{n}Y^{n}M}-P_{X^{n}Y^{n}M}\right\Vert _{TV}\leq\epsilon}\sup_{f,g}\rho_{Q}(f(X^{n},M);g(Y^{n},M)|M)\leq\beta$
for any $\epsilon>0$).
\item Privacy protection in data collection or data mining: In data collection
or data mining, privacy protection of users' data is an important
problem. To that end, we need first identify which part is common
information and which part is private information. Our result gives
a better answer to this question and hence it can be directly applied
to privacy protection in data collection or data mining.
\item Privacy constrained source simulation: As stated in \cite{Kamath},
the private sources simulation problem has natural applications in
numerous areas \textendash{} from game-theoretic coordination in a
network  to control of a dynamical system over a distributed network
with privacy protection. Our results are expected to be exploited
in many future remote-controlled applications, such as drone-based
delivery system,  privacy-preserving navigation, secure network service,
etc.
\end{itemize}

\appendices{}

\section{\label{sec:Proof-of-Equation}Proof of Equation \eqref{eq:-76}}

First we prove $\inf_{P_{U|XY}:C_{GK}(X;Y|U)=0}I(XY;U)\leq C_{GK}(X;Y)$.
Assume $f^{*},g^{*}$ achieve the supremum in \eqref{eq:-77}, then
we claim that setting $U=f^{*}\left(X\right)=g^{*}\left(Y\right)$,
it holds that $C_{GK}(X;Y|U)=0$. We use contradiction to prove this
claim. Suppose $C_{GK}(X;Y|U)>0$, i.e., there exists a pair of $f',g'$
such that $f'\left(X,U\right)=g'\left(Y,U\right)$ and $H(f'\left(X,U\right)|U)>0$.
Since $U$ is a function of $X$ and also a function of $Y$, we can
express $f'\left(X,U\right)$ as $f''\left(X\right)$ and $g'\left(Y,U\right)$
as $g''\left(Y\right)$ for some functions $f''$ and $g''$. Setting
$f\left(X\right)=\left(f^{*}\left(X\right),f''\left(X\right)\right)$
and $g\left(Y\right)=\left(g^{*}\left(Y\right),g''\left(Y\right)\right)$,
we have $f\left(X\right)=g\left(Y\right)$ and
\begin{equation}
H\left(f\left(X\right)\right)=H\left(U\right)+H(f'\left(X,U\right)|U)>H\left(U\right).
\end{equation}
This contradicts with the assumption of $f^{*},g^{*}$ achieving the
supremum in \eqref{eq:-77}. Therefore, $C_{GK}(X;Y|U)=0$. This implies
\begin{align}
{\displaystyle \inf_{P_{U|XY}:C_{GK}(X;Y|U)=0}I(XY;U)} & \leq H\left(U\right)=C_{GK}(X;Y).
\end{align}

Next we prove $\inf_{P_{U|XY}:C_{GK}(X;Y|U)=0}I(XY;U)\geq C_{GK}(X;Y)$.
We also assume $f^{*},g^{*}$ achieve the supremum in \eqref{eq:-77}.
Then we claim that for any $U$ such that $C_{GK}(X;Y|U)=0$, it holds
that $f^{*}\left(X\right)=g^{*}\left(Y\right)=\kappa\left(U\right)$
for some function $\kappa$, i.e., $U$ contains the common randomness
of $X,Y$. Next we prove this claim.

Assume $f',g'$ achieve the supremum in \eqref{eq:-78}. Then we have
$C_{GK}(X;Y|U)=0$ implies $H(f'\left(X,U\right)|U)=0$, which further
implies $f'\left(X,U\right)$ is a function of $U$; see \cite[Problem 2.5]{Cover}.
Setting $f\left(X,U\right)=\left(f^{*}\left(X\right),f'\left(X,U\right)\right)$
and $g\left(Y,U\right)=\left(g^{*}\left(Y\right),g'\left(Y,U\right)\right)$,
we have $f\left(X,U\right)=g\left(Y,U\right)$ and
\begin{equation}
H(f'\left(X,U\right)|U)\leq H(f\left(X,U\right)|U).\label{eq:-79}
\end{equation}
Owing to the optimality of $f',g'$, the equality in \eqref{eq:-79}
should hold. Therefore, $H(f^{*}\left(X\right)|U,f'\left(X,U\right))=0$.
This implies $f^{*}\left(X\right)$ is a function of $U$ and $f'\left(X,U\right)$.
Combining it with that $f'\left(X,U\right)$ is a function of $U$,
we have $f^{*}\left(X\right)$ is a function of $U$. Therefore, $f^{*}\left(X\right)=g^{*}\left(Y\right)=\kappa\left(U\right)$
for some function $\kappa$.

Using the claim, we have
\begin{align}
{\displaystyle \inf_{P_{U|XY}:C_{GK}(X;Y|U)=0}I(XY;U)} & \geq{\displaystyle \inf_{P_{U|XY}:C_{GK}(X;Y|U)=0}I(XY;\kappa\left(U\right))}\\
 & =\inf_{P_{U|XY}:C_{GK}(X;Y|U)=0}I(XY;f^{*}\left(X\right))\\
 & =H\left(f^{*}\left(X\right)\right)\\
 & =C_{GK}(X;Y).
\end{align}

Combining these two cases above, we have $\inf_{P_{U|XY}:C_{GK}(X;Y|U)=0}I(XY;U)=C_{GK}(X;Y)$.

\section{\label{sec:Proof-of-Lemma-sing}Proof of Lemma \ref{lem:Singular-value-characterization}}

A proof for the unconditional version of the lemma can be found in
\cite{Anantharam}. Here we extend the proof to the conditional version.
To that end, we only consider finite valued random variables. For
countably infinitely valued or continuous random variables, the result
can be proven similarly.

For finite valued random variables, we will show maximal correlation
$\rho_{m}(X;Y|U)$ can also be characterized by the second largest
singular value of the matrix $Q_{u}$ with entries $Q_{u}(x,y):={\displaystyle \frac{p(x,y|u)}{\sqrt{p(x|u)p(y|u)}}=\frac{p(x,y,u)}{\sqrt{p(x,u)p(y,u)}}.}$
Without loss of generality, we can rewrite
\begin{equation}
{\displaystyle \rho_{m}(X;Y|U)=\sup_{f,g}\mathbb{E}[f(X,U)g(Y,U)]},\label{eq:-75}
\end{equation}
where the maximization is taken over all $f,g$ such that $\mathbb{E}[f(X,U)]=\mathbb{E}[g(Y,U)]=0$,
$\mathbb{E}\textrm{var}(f(X,U))=\mathbb{E}\textrm{var}(g(Y,U))=1$.
Observe that
\begin{equation}
\mathbb{E}[f(X,U)g(Y,U)]={\displaystyle \sum_{x,y,u}(f(x,u)\sqrt{p(x,u)})Q_{u}(x,y)(g(y,u)\sqrt{p(y,u)})},
\end{equation}
\begin{equation}
\sum_{x}\sqrt{p(x,u)}Q_{u}(x,y)=\sqrt{p(y,u)},{\displaystyle \sum_{y}Q_{u}(x,y)\sqrt{p(y,u)}=\sqrt{p(x,u)}},
\end{equation}
and the conditions $\mathbb{E}[f(X,U)]=0$ and $\mathbb{E}[g(Y,U)]=0$
are respectively equivalent to requiring that $(x,u)\mapsto f(x,u)\sqrt{p(x,u)}$
is orthogonal to $(x,u)\mapsto\sqrt{p(x,u)}$ and that $(y,u)\mapsto g(y,u)\sqrt{p(y,u)}$
is orthogonal to $(y,u)\mapsto\sqrt{p(y,u)}.$ By Singular Value Decomposition,
$Q_{u}={\displaystyle \sum_{i=1}^{n}\lambda_{u,i}a_{u,i}b_{u,i}^{T}}$,
where $\lambda_{u,1}=1,a_{u,1}=(\sqrt{p(x,u)})_{x},b_{u,1}=(\sqrt{p(y,u)})_{y}$.
Therefore,
\begin{align}
\mathbb{E}[f(X,U)g(Y,U)] & ={\displaystyle \sum_{x,y,u}(f(x,u)\sqrt{p(x,u)})Q_{u}(x,y)(g(y,u)\sqrt{p(y,u)})}\\
 & ={\displaystyle \sum_{u}f_{u}^{T}(\sum_{i=1}^{n}\lambda_{u,i}a_{u,i}b_{u,i}^{T})g_{u}}\\
 & ={\displaystyle \sum_{u}\sum_{i=2}^{n}\lambda_{u,i}c_{u,i}d_{u,i}}\\
 & {\displaystyle \leq\sum_{u}\sum_{i=2}^{n}\lambda_{u,i}\frac{c_{u,i}^{2}+d_{u,i}^{2}}{2}},\label{eq:-33}
\end{align}
where $f_{u}:=(f(x,u)\sqrt{p(x,u)})_{x},g_{u}:=(g(y,u)\sqrt{p(y,u)})_{y},c_{u,i}:=f_{u}^{T}a_{u,i},d_{u,i}:=g_{u}^{T}b_{u,i},i\geq2$.
Furthermore,

\begin{align}
 & {\displaystyle \sum_{u}\Vert f_{u}\Vert^{2}=\sum_{u}\Vert g_{u}\Vert^{2}=1},\\
 & {\displaystyle \sum_{i=2}^{n}c_{u,i}^{2}=\Vert f_{u}\Vert^{2}},\\
 & {\displaystyle \sum_{i=2}^{n}d_{u,i}^{2}=\Vert g_{u}\Vert^{2}}.
\end{align}
Hence

\begin{align}
{\displaystyle \sum_{u}\sum_{i=2}^{n}c_{u,i}^{2}=1},\\
{\displaystyle \sum_{u}\sum_{i=2}^{n}d_{u,i}^{2}=1}.
\end{align}
Combining these with \eqref{eq:-75} and \eqref{eq:-33} gives us
\begin{equation}
{\displaystyle \rho_{m}(X;Y|U)\leq\sup_{u:P(u)>0}\lambda_{u,2}}.
\end{equation}

On the other hand, it is easy to verify that the upper bound ${\displaystyle \sup_{u:P(u)>0}\lambda_{u,2}}$
can be achieved by choosing
\begin{equation}
f_{u}=\begin{cases}
a_{u,2}, & \textrm{if }u=u^{*};\\
0, & \textrm{otherwise}.
\end{cases}
\end{equation}
and
\begin{equation}
g_{u}=\begin{cases}
b_{u,2}, & \textrm{if }u=u^{*};\\
0, & \textrm{otherwise}.
\end{cases}
\end{equation}
Therefore,
\begin{equation}
{\displaystyle \rho_{m}(X;Y|U)=\sup_{u:P(u)>0}\lambda_{u,2}}.
\end{equation}

\section{\label{sec:Proof-of-Lemma-cond}Proof of Lemma \ref{lem:Conditioning-reduces-covariance}}

By the law of total covariance, we have
\begin{equation}
\mathbb{E}\textrm{cov}(X,Y|Z)=\mathbb{E}\textrm{cov}(X,Y|ZU)+\mathbb{E}_{Z}\textrm{c}\textrm{o}\textrm{v}_{U}(\mathbb{E}(X|ZU),\mathbb{E}(Y|ZU)).
\end{equation}
Hence to prove Lemma \ref{lem:Conditioning-reduces-covariance}, we
only need to show
\begin{equation}
\sqrt{\mathbb{E}\textrm{var}(X|ZU)\mathbb{E}\textrm{var}(Y|ZU)}+\mathbb{E}_{Z}\textrm{c}\textrm{o}\textrm{v}_{U}(\mathbb{E}(X|ZU),\mathbb{E}(Y|ZU))\leq\sqrt{\mathbb{E}\textrm{var}(X|Z)\mathbb{E}\textrm{var}(Y|Z)}.\label{eq:-34}
\end{equation}

To prove this, we consider
\begin{align}
 & \mathbb{E}\textrm{var}(X|ZU)\mathbb{E}\textrm{var}(Y|ZU)\nonumber \\
= & \left(\mathbb{E}\textrm{var}(X|Z)-\mathbb{E}_{Z}\textrm{var}_{U}(\mathbb{E}(X|ZU))\right)\left(\mathbb{E}\textrm{var}(Y|Z)-\mathbb{E}_{Z}\textrm{var}_{U}(\mathbb{E}(Y|ZU))\right)\label{eq:-69}\\
= & \mathbb{E}\textrm{var}(X|Z)\mathbb{E}\textrm{var}(Y|Z)-\mathbb{E}\textrm{var}(X|Z)\mathbb{E}_{Z}\textrm{var}_{U}(\mathbb{E}(Y|ZU))\nonumber \\
 & -\mathbb{E}\textrm{var}(Y|Z)\mathbb{E}_{Z}\textrm{var}_{U}(\mathbb{E}(X|ZU))+\mathbb{E}_{Z}\textrm{var}_{U}(\mathbb{E}(X|ZU))\mathbb{E}_{Z}\textrm{var}_{U}(\mathbb{E}(Y|ZU))\\
\leq & \mathbb{E}\textrm{var}(X|Z)\mathbb{E}\textrm{var}(Y|Z)-2\sqrt{\mathbb{E}\textrm{var}(X|Z)\mathbb{E}_{Z}\textrm{var}_{U}(\mathbb{E}(Y|ZU))\cdot\mathbb{E}\textrm{var}(Y|Z)\mathbb{E}_{Z}\textrm{var}_{U}(\mathbb{E}(X|ZU))}\nonumber \\
 & +\mathbb{E}_{Z}\textrm{var}_{U}(\mathbb{E}(X|ZU))\mathbb{E}_{Z}\textrm{var}_{U}(\mathbb{E}(Y|ZU))\\
= & \left(\sqrt{\mathbb{E}\textrm{var}(X|Z)\mathbb{E}\textrm{var}(Y|Z)}-\sqrt{\mathbb{E}_{Z}\textrm{var}_{U}(\mathbb{E}(X|ZU))\mathbb{E}_{Z}\textrm{var}_{U}(\mathbb{E}(Y|ZU))}\right)^{2}\label{eq:-71}
\end{align}
where \eqref{eq:-69} follows from the law of total variance
\begin{equation}
\mathbb{E}\textrm{var}(X|Z)=\mathbb{E}_{Z}\textrm{var}_{U}(X|ZU)+\mathbb{E}_{Z}\textrm{var}_{U}(\mathbb{E}(X|ZU)).\label{eq:-70}
\end{equation}

Since $\mathbb{E}_{Z}\textrm{var}_{U}(X|ZU)\geq0$, from \eqref{eq:-70},
we have
\begin{equation}
\mathbb{E}_{Z}\textrm{var}_{U}(\mathbb{E}(X|ZU))\leq\mathbb{E}\textrm{var}(X|Z).
\end{equation}
Similarly, we have
\begin{equation}
\mathbb{E}_{Z}\textrm{var}_{U}(\mathbb{E}(Y|ZU))\leq\mathbb{E}\textrm{var}(Y|Z).
\end{equation}
Therefore,
\begin{equation}
\mathbb{E}_{Z}\textrm{var}_{U}(\mathbb{E}(X|ZU))\mathbb{E}_{Z}\textrm{var}_{U}(\mathbb{E}(Y|ZU))\leq\mathbb{E}\textrm{var}(X|Z)\mathbb{E}\textrm{var}(Y|Z).\label{eq:-37}
\end{equation}

Combining \eqref{eq:-71} and \eqref{eq:-37}, we have
\begin{align}
\sqrt{\mathbb{E}\textrm{var}(X|ZU)\mathbb{E}\textrm{var}(Y|ZU)} & \leq\sqrt{\mathbb{E}\textrm{var}(X|Z)\mathbb{E}\textrm{var}(Y|Z)}-\sqrt{\mathbb{E}_{Z}\textrm{var}_{U}(\mathbb{E}(X|ZU))\mathbb{E}_{Z}\textrm{var}_{U}(\mathbb{E}(Y|ZU))}.
\end{align}
Furthermore, by Cauchy-Schwarz inequality, it holds that
\begin{align}
|\mathbb{E}_{Z}\textrm{c}\textrm{o}\textrm{v}_{U}(\mathbb{E}(X|ZU),\mathbb{E}(Y|ZU))| & =|\mathbb{E}\left[\left(\mathbb{E}(X|ZU)-\mathbb{E}(X|Z)\right)\left(\mathbb{E}(Y|ZU)-\mathbb{E}(Y|Z)\right)\right]|\label{eq:-36}\\
 & \leq\sqrt{\mathbb{E}\left(\mathbb{E}(X|ZU)-\mathbb{E}(X|Z)\right)^{2}\cdot\mathbb{E}\left(\mathbb{E}(Y|ZU)-\mathbb{E}(Y|Z)\right)^{2}}\\
 & =\sqrt{\mathbb{E}_{Z}\textrm{var}_{U}(\mathbb{E}(X|ZU))\mathbb{E}_{Z}\textrm{var}_{U}(\mathbb{E}(Y|ZU))}.
\end{align}
Therefore,
\begin{align}
\sqrt{\mathbb{E}\textrm{var}(X|ZU)\mathbb{E}\textrm{var}(Y|ZU)} & \leq\sqrt{\mathbb{E}\textrm{var}(X|Z)\mathbb{E}\textrm{var}(Y|Z)}-|\mathbb{E}_{Z}\textrm{c}\textrm{o}\textrm{v}_{U}(\mathbb{E}(X|ZU),\mathbb{E}(Y|ZU))|\\
 & \leq\sqrt{\mathbb{E}\textrm{var}(X|Z)\mathbb{E}\textrm{var}(Y|Z)}-\mathbb{E}_{Z}\textrm{c}\textrm{o}\textrm{v}_{U}(\mathbb{E}(X|ZU),\mathbb{E}(Y|ZU)),
\end{align}
which implies \eqref{eq:-34}. This completes the proof.

\section{\label{sec:Proof-of-Theorem-Gaussian}Proof of Theorem \ref{thm:Gaussian}}

From Theorem \ref{thm:Lower-bound}, the following inequality follows
immediately.
\begin{equation}
C_{\beta}^{(G)}(X;Y){\displaystyle \geq\frac{1}{2}\log^{+}\left(\frac{1+\beta_{0}}{1-\beta_{0}}/\frac{1+\beta}{1-\beta}\right)}.\label{eq:-38}
\end{equation}

On the other hand, $(X,Y)$ can be expressed as
\begin{align}
X & =\alpha U+\sqrt{1-\alpha^{2}}Z_{1},\\
Y & =\alpha U+\sqrt{1-\alpha^{2}}Z_{2},
\end{align}
with
\begin{equation}
\alpha=\sqrt{\frac{\beta_{0}-\beta}{1-\beta}}
\end{equation}
and the covariance of $(Z_{1},Z_{2})$
\begin{equation}
\Sigma_{(Z_{1},Z_{2})}=\left(\begin{array}{ll}
1 & \beta\\
\beta & 1
\end{array}\right),
\end{equation}
where $U\sim\mathcal{N}(0,1)$. Hence we have
\begin{equation}
\rho(X,Y|U)\leq\beta
\end{equation}
and
\begin{equation}
I(XY;U)={\displaystyle \frac{1}{2}\log^{+}\left(\frac{1+\beta_{0}}{1-\beta_{0}}/\frac{1+\beta}{1-\beta}\right)}.
\end{equation}
Hence
\begin{equation}
C_{\beta}^{(G)}(X;Y){\displaystyle \leq\frac{1}{2}\log^{+}\left(\frac{1+\beta_{0}}{1-\beta_{0}}/\frac{1+\beta}{1-\beta}\right)}.\label{eq:-39}
\end{equation}

Combining \eqref{eq:-38} and \eqref{eq:-39} gives us
\begin{equation}
C_{\beta}^{(G)}(X;Y)={\displaystyle \frac{1}{2}\log^{+}\left(\frac{1+\beta_{0}}{1-\beta_{0}}/\frac{1+\beta}{1-\beta}\right)}.
\end{equation}
This completes the proof.

\section{\label{sec:Proof-of-Theorem-PSS}Proof of Theorem \ref{thm:app-cent-PSS}}

\subsection{{\label{subsec:PSSAchievability}Achievability}}

\textbf{Codebook Generation:} Suppose $R>C_{\beta}(X;Y)$. Randomly
and independently generate sequences $u^{n}(m),m\in[1:2^{nR}]$ with
each according to ${\displaystyle \prod_{i=1}^{n}P_{U}(u_{i})}$.
The codebook $C=\{u^{n}(m),m\in[2^{nR}]\}$.

\textbf{Input Process Generator: }Generate input source $M$ according
to the uniform distribution over $[2^{nR}].$

\textbf{Source Generator:} Upon $m$, the generator generates sources
$(X^{n},Y^{n})$ according to ${\displaystyle \prod_{i=1}^{n}P_{XY|U}(x_{i},y_{i}|u_{i}(m))}$.

For such generator, the induced overall distribution is
\begin{equation}
P_{X^{n}Y^{n}M}(x^{n},{\displaystyle y^{n},m):=2^{-nR}\prod_{i=1}^{n}P_{XY|U}(x_{i},y_{i}|u_{i}(m))}.
\end{equation}

According to soft-cover lemma \cite{Cuff}, if $R>I(XY;U)$, then
\begin{equation}
{\displaystyle \lim_{n\rightarrow\infty}\mathbb{E}_{\mathcal{C}}\Vert P_{X^{n}Y^{n}}-\prod_{i=1}^{n}P_{XY}\Vert_{TV}=0}.
\end{equation}

Given $U^{n}(m)=u^{n},$ $(X^{n},Y^{n})$ is a conditionally independent
sequence, i.e.,
\begin{equation}
P_{X^{n}Y^{n}|M}(x^{n},{\displaystyle y^{n}|m)=\prod_{i=1}^{n}P_{XY|U}(x_{i},y_{i}|u_{i}(m))}.
\end{equation}
Hence according to Lemma \ref{lem:MCsequence}, we get
\begin{equation}
{\displaystyle \rho_{m}(X^{n};Y^{n}|M)=\sup_{1\leq i\leq n}\rho_{m}(X_{i};Y_{i}|U_{i}(M))}.
\end{equation}
Furthermore, from Lemma \ref{lem:Alternative-characterization}, we
have
\begin{equation}
{\displaystyle \rho_{m}(X_{i};Y_{i}|U_{i}(M))=\sup_{u:P_{U_{i}}(u)>0}\lambda_{2,P_{XY|U}}(u)\leq\sup_{u:P_{U}(u)>0}\lambda_{2,P_{XY|U}}(u)\leq\beta}.
\end{equation}
Hence
\begin{equation}
\rho_{m}(X^{n};Y^{n}|M)\leq\beta.
\end{equation}
This implies
\begin{equation}
R_{PSS}(\beta)\leq C_{\beta}(X;Y).
\end{equation}

\subsection{{Converse}}

Assume there exists a sequence of distributed generators such that
$\limsup_{n\rightarrow\infty}\frac{1}{n}H(M)\leq R,\,\rho_{m}(X^{n};Y^{n}|M)\leq\beta,\forall n,$
and $\lim_{n\rightarrow\infty}\Vert P_{X^{n}Y^{n}}-\prod P_{XY}\Vert_{TV}=0$.
Consider that
\begin{align}
\frac{1}{n}I(X^{n}Y^{n};M) & ={\displaystyle \frac{1}{n}\sum_{i=1}^{n}I(X_{i}Y_{i};M|X^{i-1}Y^{i-1})}\\
 & ={\displaystyle \frac{1}{n}\sum_{i=1}^{n}H(X_{i}Y_{i}|X^{i-1}Y^{i-1})-H(X_{i}Y_{i}|MX^{i-1}Y^{i-1})}\\
 & ={\displaystyle \frac{1}{n}\sum_{i=1}^{n}H_{Q}(X_{i}Y_{i})-H(X_{i}Y_{i}|MX^{i-1}Y^{i-1})}\\
 & ={\displaystyle \frac{1}{n}\sum_{i=1}^{n}H(X_{i}Y_{i})-H(X_{i}Y_{i}|MX^{i-1}Y^{i-1})}\\
 & ={\displaystyle \frac{1}{n}\sum_{i=1}^{n}I(X_{i}Y_{i};MX^{i-1}Y^{i-1})}\\
 & =I(X_{T}Y_{T};MX^{T-1}Y^{T-1}|T)\\
 & =I(X_{T}Y_{T};MX^{T-1}Y^{T-1}T)\\
 & \geq I(X_{T}Y_{T};MT)\\
 & =I(XY;V),
\end{align}
where $T$ is a time-sharing random variable uniformly distributed
$[1:n]$ and independent of all other random variables, and $X:=X_{T},Y:=Y_{T},V:=MT$.
Combining the inequality above with
\begin{equation}
{\displaystyle \frac{1}{n}I(X^{n}Y^{n};M)\leq\frac{1}{n}H(M)\leq R}
\end{equation}
gives us
\begin{equation}
I(XY;V)\leq R.\label{eq:-42}
\end{equation}
On the other hand,
\begin{align}
\rho_{m}(X^{n};Y^{n}|M) & {\displaystyle \geq\sup_{i}\rho_{m}(X_{i};Y_{i}|M)}\label{eq:-40}\\
 & ={\displaystyle \sup_{i,m}\rho_{m}(X_{i};Y_{i}|M=m)}\label{eq:-41}\\
 & ={\displaystyle \sup_{i,m}\rho_{m}(X_{T};Y_{T}|M=m,T=i)}\\
 & =\rho_{m}(X_{T};Y_{T}|M,T)\\
 & =\rho_{m}(X;Y|V),\label{eq:-43}
\end{align}
where \eqref{eq:-40} follows from the definition of maximal correlation,
and \eqref{eq:-41} follows from Lemma \ref{lem:Alternative-characterization}.

Combining \eqref{eq:-42} with \eqref{eq:-43} gives us
\begin{equation}
R{\displaystyle \geq\inf_{P_{U|X,Y}:\rho_{m}(X;Y|V)\leq\beta}I(X,Y;V)=C_{\beta}(X;Y)}.
\end{equation}
Hence
\begin{equation}
R_{PSS}(\beta)\geq C_{\beta}(X;Y).
\end{equation}
This completes the proof.

\section{\label{sec:Proof-of-Theorem-CIE}Proof of Theorem \ref{thm:app-cent-CIE}}

\subsection{{Achievability}}

\textbf{Codebook Generation:} Suppose $R>C_{\beta}(X;Y)$. Randomly
and independently generate sequences $u^{n}(m),m\in[1:2^{nR}]$ with
each according to ${\displaystyle \prod_{i=1}^{n}P_{U}(u_{i})}$.
The codebook $C=\{u^{n}(m),m\in[2^{nR}]\}$.

\textbf{Extractor:} Upon $(X,Y^{n})$, the extractor generates sources
$m$ using a likelihood encoder $P_{M|X^{n}Y^{n}}(m|x^{n},y^{n})\propto{\displaystyle \prod_{i=1}^{n}P_{XY|U}(x_{i},y_{i}|u_{i}(m))}$,
where $\propto$indicates that appropriate normalization is required.

For such extractor, the induced overall distribution $P_{X^{n}Y^{n}M}$
is related to an ideal distribution
\begin{equation}
Q_{X^{n}Y^{n}M}(x^{n},y^{n},m):=2^{-nR}{\displaystyle \prod_{i=1}^{n}P_{XY|U}(x_{i},y_{i}|u_{i}(m))}.
\end{equation}
According to soft-covering lemma \cite{Cuff}, if $R>I(XY;U)$, then
\begin{equation}
{\displaystyle \lim_{n\rightarrow\infty}\mathbb{E}_{\mathcal{C}}\Vert P_{X^{n}Y^{n}}-Q_{X^{n}Y^{n}}\Vert_{TV}=0},
\end{equation}
where
\begin{equation}
Q_{X^{n}Y^{n}}(x^{n},{\displaystyle y^{n})=\prod_{i=1}^{n}P_{XY}(x_{i},y_{i})}.
\end{equation}
On the other hand, observe that $P_{M|X^{n}Y^{n}}=Q_{M|X^{n}Y^{n}}$.
Hence by Property \ref{pr:properties}, we further have
\begin{equation}
{\displaystyle \lim_{n\rightarrow\infty}\mathbb{E}_{\mathcal{C}}\Vert P_{X^{n}Y^{n}M}-Q_{X^{n}Y^{n}M}\Vert_{TV}=\lim_{n\rightarrow\infty}\mathbb{E}_{\mathcal{C}}\Vert P_{X^{n}Y^{n}}-Q_{X^{n}Y^{n}}\Vert_{TV}=0}.
\end{equation}

Given $U^{n}(m)=u^{n},$ $(X^{n}Y^{n})$ is an independently distributed
sequence under distribution $Q$. That is
\begin{equation}
Q_{X^{n}Y^{n}|M}(x^{n},{\displaystyle y^{n}|m)=\prod_{i=1}^{n}P_{XY|U}(x_{i},y_{i}|u_{i}(m))}.
\end{equation}
Hence according to Lemma \ref{lem:MCsequence}, we get
\begin{equation}
{\displaystyle \rho_{m,Q}(X^{n};Y^{n}|M)=\sup_{1\leq i\leq n}\rho_{m,Q}(X_{i};Y_{i}|U_{i}(M))}.
\end{equation}
Furthermore, from Lemma \ref{lem:Alternative-characterization}, we
have
\begin{equation}
{\displaystyle \rho_{m,Q}(X_{i};Y_{i}|U_{i}(M))=\sup_{u:P_{U_{i}}(u)>0}\lambda_{2,P_{XY|U}}(u)\leq\sup_{u:P_{U}(u)>0}\lambda_{2,P_{XY|U}}(u)\leq\beta}.
\end{equation}
Hence
\begin{equation}
\rho_{m,Q}(X^{n};Y^{n}|M)\leq\beta.
\end{equation}
This implies
\begin{equation}
R_{CIE}(\beta)\leq C_{\beta}(X;Y).
\end{equation}

\subsection{\label{subsec:Converse}{Converse}}

Assume there exists a sequence of extractors such that
\begin{equation}
\limsup_{n\rightarrow\infty}\frac{1}{n}H(M)\leq R,\label{eq:-30-1-1-1}
\end{equation}
and
\begin{equation}
{\displaystyle \inf_{Q_{X^{n},Y^{n},M}:\Vert Q_{X^{n},Y^{n},M}-P_{X^{n},Y^{n},M}\Vert_{TV}\leq\epsilon_{n}}\rho_{m,Q}(X^{n};Y^{n}|M)\leq\beta},\forall n,\label{eq:-67}
\end{equation}
for some $\epsilon_{n}$ such that $\lim_{n\rightarrow\infty}\epsilon_{n}=0$.

Assume $Q_{X^{n},Y^{n},M}$ achieves the infimum in \eqref{eq:-67}.
Hence $\Vert Q_{X^{n},Y^{n},M}-P_{X^{n},Y^{n},M}\Vert_{TV}\rightarrow0$.
Then by the total-variation bound on entropy, we have
\begin{align}
 & |{\displaystyle \frac{1}{n}H_{P}(X^{n}Y^{n}M)-\frac{1}{n}H_{Q}(X^{n}Y^{n}M)|}\nonumber \\
 & {\displaystyle \leq\frac{1}{n}2\Vert Q_{X^{n},Y^{n},M}-P_{X^{n},Y^{n},M}\Vert_{TV}\log\frac{|\mathcal{X}^{n}\times\mathcal{Y}^{n}\times[2^{nR}]|}{2\Vert Q_{X^{n},Y^{n},M}-P_{X^{n},Y^{n}M}\Vert_{TV}}}\\
 & =2{\displaystyle \Vert Q_{X^{n},Y^{n},M}-P_{X^{n},Y^{n},M}\Vert_{TV}\log\frac{2^{R}|\mathcal{X}||\mathcal{Y}|}{2\Vert Q_{X^{n},Y^{n},M}-P_{X^{n},Y^{n},M}\Vert_{TV}}}\\
 & \rightarrow0,
\end{align}
and similarly,
\begin{equation}
|{\displaystyle \frac{1}{n}H_{P}(X^{n}Y^{n})-\frac{1}{n}H_{Q}(X^{n}Y^{n})|\leq2\Vert Q_{X^{n},Y^{n}}-P_{X^{n},Y^{n}}\Vert_{TV}\log\frac{|\mathcal{X}||\mathcal{Y}|}{2\Vert Q_{X^{n},Y^{n}}-P_{X^{n},Y^{n}}\Vert_{TV}}\rightarrow0},
\end{equation}
and
\begin{equation}
|{\displaystyle \frac{1}{n}H_{P}(M)-\frac{1}{n}H_{Q}(M)|\leq2\Vert Q_{M}-P_{M}\Vert_{TV}\log\frac{2^{R}}{2\Vert Q_{M}-P_{M}\Vert_{TV}}\rightarrow0}.
\end{equation}

Furthermore, observe ${\displaystyle \frac{1}{n}I(X^{n}Y^{n};M)=\frac{1}{n}H(X^{n}Y^{n})+\frac{1}{n}H(M)-\frac{1}{n}H(X^{n}Y^{n}M)}$.
Hence
\begin{align}
\frac{1}{n}I_{P}(X^{n}Y^{n};M) & {\displaystyle \leq\frac{1}{n}H_{P}(M)}\\
 & \leq R.\label{eq:-45}
\end{align}
On the other hand, consider that
\begin{align}
I_{P}(X^{n}Y^{n};M) & ={\displaystyle \sum_{i=1}^{n}I_{P}(X_{i}Y_{i};M|X^{i-1}Y^{i-1})}\\
 & ={\displaystyle \sum_{i=1}^{n}I_{P}(X_{i}Y_{i};MX^{i-1}Y^{i-1})}\\
 & =nI_{P}(X_{T}Y_{T};MX^{T-1}Y^{T-1}|T)\\
 & =nI_{P}(X_{T}Y_{T};MX^{T-1}Y^{T-1}T)\\
 & \geq nI_{P}(X_{T}Y_{T};MT)\\
 & \geq nI_{Q}(X_{T}Y_{T};MT)-n\epsilon_{n}\\
 & =nI_{Q}(XY;V)-n\epsilon_{n},
\end{align}
where $T$ is a time-sharing random variable uniformly distributed
$[1:n]$ and independent of all other random variables, and $X:=X_{T},Y:=Y_{T},V:=MT$.
Combining the inequality above with \eqref{eq:-45} gives us
\begin{equation}
I_{Q}(XY;V)\leq R+\epsilon_{n}.\label{eq:-49}
\end{equation}
Furthermore,
\begin{align}
\rho_{m,Q}(X^{n};Y^{n}|M) & {\displaystyle \geq\max_{i}\rho_{m,Q}(X_{i};Y_{i}|M)}\label{eq:-44}\\
 & ={\displaystyle \max\rho_{m,Q}(X_{i};Y_{i}i,m|M=m)}\label{eq:-46}\\
 & ={\displaystyle \max\rho_{m,Q}(X_{T};Y_{T}i,m|M=m,T=i)}\\
 & =\rho_{m,Q}(X_{T};Y_{T}|M,T)\label{eq:-47}\\
 & =\rho_{m,Q}(X;Y|V),
\end{align}
where \eqref{eq:-44} follows from the definition of maximal correlation,
and \eqref{eq:-46} and \eqref{eq:-47} follow from Lemma \ref{lem:Alternative-characterization}.
Furthermore, \eqref{eq:-67} implies $\limsup_{n\rightarrow\infty}{\displaystyle \rho_{m,Q}(X^{n};Y^{n}|M)\leq\beta}$.
Hence
\begin{equation}
\rho_{m,Q}(X;Y|V)\leq\beta.\label{eq:-50}
\end{equation}

Combining \eqref{eq:-49} with \eqref{eq:-50} gives us
\begin{equation}
R\geq{\displaystyle \inf_{P_{V|X,Y}:\rho_{m}(X;Y|V)\leq\beta}I(XY;V)-\epsilon_{n}=C_{\beta}(X;Y)-\epsilon_{n}}.
\end{equation}
Hence
\begin{equation}
R_{CIE}(\beta)\geq C_{\beta}(X;Y).
\end{equation}
This completes the proof.

\section{\label{sec:Proof-of-Theorem-CIE-dist}Proof of Theorem \ref{thm:app-dist-CIE}}

\subsection{{Achievability}}

For the achievability part, we only need to show the upper bound $C_{\beta}^{(D,UB)}(X;Y)$
is achievable. It is also equivalent to showing that $\left(R,\beta\right)$
with $R>C_{\beta}^{(D,UB)}(X;Y)$ is achievable. Next we use a random
binning strategy, OSRB (Output Statistics of Random Binning) \cite{Yassaee}
to prove this, instead of using soft-covering technique. This is because
the \textquotedblleft soft-covering\textquotedblright{} lemma is not
easily applicable to complicated network structures, but OSRB is.
Furthermore, it is worth noting that the random binning technique
can be applied to prove the centralized setting case as well. Next
we give the proof by following the basic proof steps of \cite{Yassaee}.

\begin{figure}
\centering\includegraphics[width=0.45\linewidth]{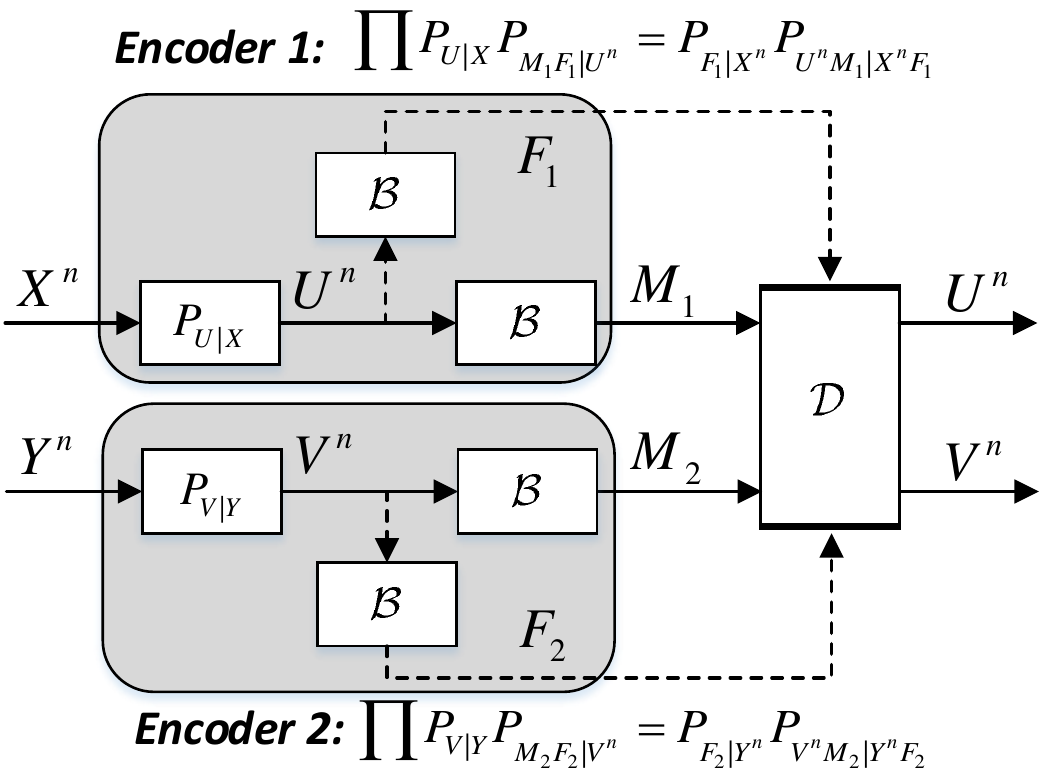}\qquad{}\includegraphics[width=0.45\linewidth]{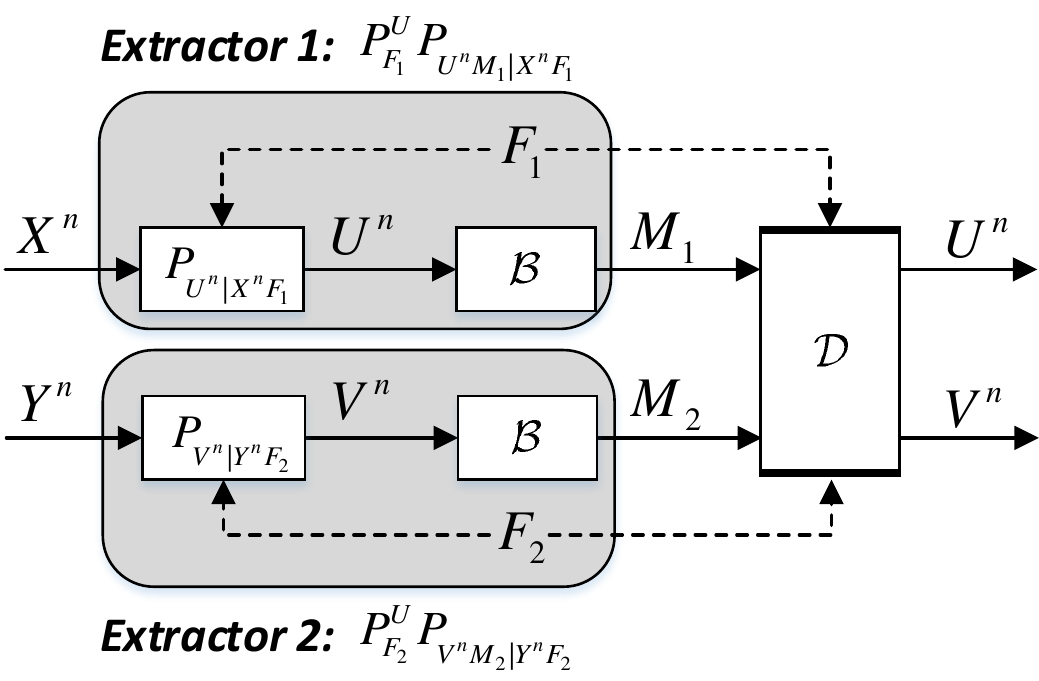}
\caption{\label{fig:RD} {\small{}(Left) Source coding side of the problem
(Protocol A). We pass i.i.d. sources $X^{n}$ and $Y^{n}$ through
virtual discrete memoryless channels $P_{U|X}$ and $P_{V|Y}$ respectively
to generate i.i.d. sequences $U^{n}$ and $V^{n}$. We describe $U^{n}$
and $V^{n}$ through two random bins $M_{i}$ and $F_{i}$ at rates
$R_{i}$ and $\tilde{R}_{i}$, $i=1,2$, where $M_{i}$ will serve
as the message for the receiver $i$ in the main problem, while $F_{i}$
will serve as the shared randomness. We use SW decoder for decoding.
(Right) The common information extraction problem assisted with the
shared randomness (Protocol B). We pass the sources $X^{n}$ and $Y^{n}$
and the shared randomnesses $F_{1}$ and $F_{2}$ through the reverse
encoders to generate sequences $U^{n}$ and $V^{n}$.  The joint
distribution of $X^{n},Y^{n},M_{1},M_{2},F_{1},F_{2}$ of protocol
A is equal to that of protocol B in total variation sense. }}
\end{figure}

\emph{Part (1) of the proof:} We define two protocols, source coding
side of the problem (Protocol A) and the main problem (Protocol B).
Fig. \ref{fig:RD} illustrates how the source coding side of the problem
can be used to prove the common information extraction problem.

\textbf{Protocol A}\emph{ (Source coding side of the problem).} Let
$(X^{n},Y^{n},U^{n},V^{n})$ be i.i.d and distributed according to
$P_{XY}P_{U|X}P_{V|Y}$. Consider the following random binning (see
the left diagram of Fig. \ref{fig:RD}): uniformly and independently
assign two bin indices $m_{1}\in[1:2^{nR_{1}}]$ and $f_{1}\in[1:2^{n\tR_{1}}]$
to each sequence $u^{n}$; and similarly, uniformly and independently
assign two bin indices $m_{2}\in[1:2^{nR_{2}}]$ and $f_{2}\in[1:2^{n\tR_{2}}]$
to each sequence $v^{n}$. Furthermore, we use Slepian-Wolf (SW) decoders
to recover $u^{n},v^{n}$ from $(m_{1},m_{2},f_{1},f_{2})$. Denote
the outputs of the decoders by $\hat{u}^{n}$ and $\hat{v}^{n}$,
respectively.

The pmf induced by the random binning, denoted by $P$, can be expressed
as
\begin{align}
 & P(x^{n},y^{n},u^{n},v^{n},f_{1},f_{2},m_{1},m_{2},\hat{u}^{n},\hat{v}^{n})\nonumber \\
 & =P(x^{n},y^{n})P(u^{n}|x^{n})P(v^{n}|y^{n})P(f_{1}|u^{n})P(f_{2}|v^{n})P(m_{1}|u^{n})P(m_{2}|v^{n})P^{SW}(\hat{u}^{n},\hat{v}^{n}|m_{1},m_{2},f_{1},f_{2})\\
 & =P(x^{n},y^{n})P(f_{1},u^{n}|x^{n})P(f_{2},v^{n}|y^{n})P(m_{1}|u^{n})P(m_{2}|v^{n})P^{SW}(\hat{u}^{n},\hat{v}^{n}|m_{1},m_{2},f_{1},f_{2})\\
 & =P(x^{n},y^{n})P(f_{1}|x^{n})P(f_{2}|y^{n})P(u^{n}|x^{n},f_{1})P(v^{n}|y^{n},f_{2})P(m_{1}|u^{n})P(m_{2}|v^{n})P^{SW}(\hat{u}^{n},\hat{v}^{n}|m_{1},m_{2},f_{1},f_{2}).\label{eq:pmfL0}
\end{align}

\textbf{Protocol B }\emph{(Common information extraction problem assisted
with the shared randomness). } In this protocol we assume that the
transmitters (extractors) and the receivers have access to the shared
randomnesses $F_{1},F_{2}$ where $F_{i}$ is uniformly distributed
over $[1:2^{n\tR_{i}}]$, $i=1,2$. Then, the protocol proceeds as
follows (see also the right diagram of Fig. \ref{fig:RD}):
\begin{itemize}
\item The transmitter 1 generates $U^{n}$ according to the conditional
pmf $P(u^{n}|x^{n},f_{1})$ of protocol A; and the transmitter 2 generates
$V^{n}$ according to the conditional pmf $P(v^{n}|y^{n},f_{2})$
of protocol A.
\item Next, knowing $u^{n}$, the transmitter 1 generates $m_{1}$ according
to the conditional pmf $P(m_{1}|u^{n})$ of protocol A. Similarly,
the transmitter 2 generates $m_{2}$ according to the conditional
pmf $P(m_{2}|v^{n})$ of protocol A.
\item Finally, upon $(m_{1},m_{2},f_{1},f_{2})$, the receiver uses the
Slepian-Wolf decoder $P^{SW}(\hat{u}^{n},\hat{v}^{n}|m_{1},m_{2},f_{1},f_{2})$
of protocol A to obtain an estimate of $\left(u^{n},v^{n}\right)$.
\end{itemize}
The pmf induced by the protocol, denoted by $\widetilde{P}$, can
be expressed as
\begin{align}
 & \widetilde{P}(x^{n},y^{n},u^{n},v^{n},f_{1},f_{2},m_{1},m_{2},\hat{u}^{n},\hat{v}^{n})\nonumber \\
 & =P(x^{n},y^{n})P^{U}(f_{1})P^{U}(f_{2})P(u^{n}|x^{n},f_{1})P(v^{n}|y^{n},f_{2})P(m_{1}|u^{n})P(m_{2}|v^{n})P^{SW}(\hat{u}^{n},\hat{v}^{n}|m_{1},m_{2},f_{1},f_{2}).\label{eq:pmfL2}
\end{align}

\emph{Part (2a) of the proof (Sufficient conditions that make the
induced pmfs approximately the same}):  Observe that $f_{1}$ is
a bin index of $u^{n}$ and $f_{2}$ is a bin index of $v^{n}$ in
protocol A. For the random binning in protocol A, \cite[Thm. 1]{Yassaee}
says that if
\begin{align}
\tR_{1} & <H(U|XY)\\
\tR_{2} & <H(V|XY)\\
\tR_{1}+\tR_{2} & <H(UV|XY)
\end{align}
then $P(x^{n},y^{n})P(f_{1}|x^{n})P(f_{2}|y^{n})\apx{}P(x^{n},y^{n})P^{U}(f_{1})P^{U}(f_{2})$.
Combining this with \eqref{eq:pmfL0} and \eqref{eq:pmfL2} gives
us
\begin{align}
\widetilde{P}(x^{n},y^{n},u^{n},v^{n},f_{1},f_{2},m_{1},m_{2},\hat{u}^{n},\hat{v}^{n}) & \apx{}P(x^{n},y^{n},u^{n},v^{n},f_{1},f_{2},m_{1},m_{2},\hat{u}^{n},\hat{v}^{n}).\label{eq:-52}
\end{align}

\emph{Part (2b) of the proof (Sufficient conditions that make the
Slepian-Wolf decoders succeed}):  \cite[Lem. 1]{Yassaee} says that
if
\begin{align}
R_{1}+\tR_{1} & >H(U|V)\\
R_{2}+\tR_{2} & >H(V|U)\\
R_{1}+R_{2}+\tR_{1}+\tR_{2} & >H(UV)
\end{align}
then
\begin{equation}
P(x^{n},y^{n},u^{n},v^{n},f_{1},f_{2},m_{1},m_{2},\hat{u}^{n},\hat{v}^{n})\apx{}P(x^{n},y^{n},u^{n},v^{n},f_{1},f_{2},m_{1},m_{2})1\{\hat{u}^{n}=u^{n},\hat{v}^{n}=v^{n}\}.\label{eq:Lpmf1.5}
\end{equation}

Using \eqref{eq:-52}, \eqref{eq:Lpmf1.5} and the triangle inequality,
we have
\begin{align}
\widetilde{P}(x^{n},y^{n},u^{n},v^{n},f_{1},f_{2},m_{1},m_{2},\hat{u}^{n},\hat{v}^{n}) & \apx{}P(x^{n},y^{n},u^{n},v^{n},f_{1},f_{2},m_{1},m_{2})1\{\hat{u}^{n}=u^{n},\hat{v}^{n}=v^{n}\}.\label{eq:Lpmf1}
\end{align}

\emph{Part (3) of the proof (Eliminating the shared randomness }$F_{1},F_{2}$\emph{):}
\eqref{eq:Lpmf1} holds for the random pmfs induced by random binning,
by Property \ref{pr:properties}, which guarantees existence of a
fixed binning such that \eqref{eq:Lpmf1} holds for the induced non-random
pmfs. \eqref{eq:Lpmf1} can be rewritten as
\begin{align}
 & \widetilde{P}(x^{n},y^{n},u^{n},v^{n},f_{1},f_{2},m_{1},m_{2},\hat{u}^{n},\hat{v}^{n})\nonumber \\
 & \apx{}P(f_{1},f_{2},m_{1},m_{2},\hat{u}^{n},\hat{v}^{n})P(x^{n},y^{n}|u^{n},v^{n})1\{\hat{u}^{n}=u^{n},\hat{v}^{n}=v^{n}\}.\label{eq:-53}
\end{align}

From \eqref{eq:-53} we further have
\begin{align}
 & \widetilde{P}(x^{n},y^{n},f_{1},f_{2},m_{1},m_{2},\hat{u}^{n},\hat{v}^{n})\nonumber \\
 & \apx{}P(f_{1},f_{2},m_{1},m_{2},\hat{u}^{n},\hat{v}^{n})P_{X^{n}Y^{n}|U^{n}V^{n}}(x^{n},y^{n}|\hat{u}^{n},\hat{v}^{n})\\
 & =P(f_{1},f_{2},m_{1},m_{2})1\left\{ \hat{u}^{n}=\hat{u}^{n}\left(m_{1},f_{1}\right),\hat{v}^{n}=\hat{v}^{n}\left(m_{2},f_{2}\right)\right\} P_{X^{n}Y^{n}|U^{n}V^{n}}(x^{n},y^{n}|\hat{u}^{n},\hat{v}^{n}),\label{eq:-54}
\end{align}
where $P_{X^{n}Y^{n}|U^{n}V^{n}}=\prod_{i=1}^{n}P_{XY|UV}$, and $\hat{u}^{n}\left(m_{1},f_{1}\right)$
and $\hat{v}^{n}\left(m_{2},f_{2}\right)$ correspond to the Slepian-Wolf
decoders. Hence
\begin{align}
 & \widetilde{P}(x^{n},y^{n},f_{1},f_{2},m_{1},m_{2})\nonumber \\
 & \apx{}Q(x^{n},y^{n},f_{1},f_{2},m_{1},m_{2})\\
 & :=P(f_{1},f_{2},m_{1},m_{2})P_{X^{n}Y^{n}|U^{n}V^{n}}(x^{n},y^{n}|\hat{u}^{n}\left(m_{1},f_{1}\right),\hat{v}^{n}\left(m_{2},f_{2}\right)).\label{eq:-55}
\end{align}

Observe that under $Q$, given $F_{1}F_{2}M_{1}M_{2}$, $X^{n}Y^{n}$
follows

\begin{equation}
Q_{X^{n}Y^{n}|F_{1}F_{2}M_{1}M_{2}}(x^{n},y^{n}|f_{1},f_{2},m_{1},m_{2})=\prod_{i=1}^{n}P_{XY|UV}(x_{i},y_{i}|\hat{u}_{i}\left(m_{1},f_{1}\right),\hat{v}_{i}\left(m_{2},f_{2}\right)).
\end{equation}
Hence by Lemma \ref{lem:MCsequence}, we get
\begin{equation}
{\displaystyle \rho_{m,Q}(X^{n};Y^{n}|F_{1}F_{2}M_{1}M_{2})=\sup_{1\leq i\leq n}\rho_{m,Q}(X_{i};Y_{i}|\hat{U}_{i}\left(M_{1},F_{1}\right),\hat{V}_{i}\left(M_{2},F_{2}\right))}.
\end{equation}
On the other hand, from Lemma \ref{lem:Alternative-characterization},
we have
\begin{align}
\rho_{m,Q}(X_{i};Y_{i}|\hat{U}_{i}\left(M_{1},F_{1}\right),\hat{V}_{i}\left(M_{2},F_{2}\right)) & =\sup_{u,v:P_{U_{i}V_{i}}(u,v)>0}\lambda_{2,P_{XY|UV}}(u,v)\\
 & \leq\sup_{u,v:P_{UV}(u,v)>0}\lambda_{2,P_{XY|UV}}(u,v)\\
 & \leq\beta.
\end{align}
Therefore,
\begin{equation}
\rho_{m,Q}(X^{n};Y^{n}|F_{1}F_{2}M_{1}M_{2})\leq\beta.
\end{equation}
By choosing $F_{1}=f_{1},F_{2}=f_{2}$ for arbitrary $\left(f_{1},f_{2}\right)$,
it holds that
\begin{equation}
\rho_{m,Q}(X^{n};Y^{n}|F_{1}=f_{1},F_{2}=f_{2},M_{1},M_{2})\leq\beta.
\end{equation}

Finally, specifying $P(m_{1}|x^{n},f_{1})$ as the encoder 1 and $P(m_{2}|x^{n},f_{2})$
as the encoder 2 (which is equivalent to, for encoder 1, generating
random sequences $u^{n}$ according to $P(u^{n}|x^{n},f_{1})$ and
then transmitting the bin index $m_{1}$ assigned to $u^{n}$, and
for encoder 2, doing similar operations), and $P^{SW}(\hat{u}^{n},\hat{v}^{n}|m_{1},m_{2},f_{1},f_{2})$
as the decoder results in a pair of encoder-decoder obeying the desired
constraints:
\begin{align}
\rho_{m,Q}(X^{n};Y^{n}|M_{1}M_{2})=\rho_{m,Q}(X^{n};Y^{n}|F_{1}=f_{1},F_{2}=f_{2},M_{1},M_{2}) & \leq\beta.
\end{align}

Observe that the common information extraction above only requires
$R_{1}+R_{2}>I(XY;UV)=I_{Q}(XY;UV)$. This implies $C_{\beta}^{(D,UB)}(X;Y)$
is achievable, which in turn implies
\begin{equation}
R_{DCIE}(\beta)\leq C_{\beta}^{(D,UB)}(X;Y).
\end{equation}

Furthermore, $C_{\beta}^{(D)}(X;Y)=\lim_{n\rightarrow\infty}\inf_{P_{U|X^{n}}P_{V|Y^{n}}:\rho_{m}(X^{n};Y^{n}|UV)\leq\beta}\frac{1}{n}{\displaystyle I(X^{n}Y^{n};UV)}$
is also achievable, since it is a multiletter extension of $C_{\beta}^{(D,UB)}(X;Y)$.

\subsection{{Converse}}

Assume there exists an extractor such that
\begin{equation}
{\displaystyle \inf_{Q_{X^{n},Y^{n},M_{1},M_{2}}:\Vert Q_{X^{n},Y^{n},M_{1},M_{2}}-P_{X^{n},Y^{n},M_{1},M_{2}}\Vert_{TV}\leq\epsilon_{n}}\rho_{m,Q}(X^{n};Y^{n}|M_{1},M_{2})\leq\beta},\forall n,
\end{equation}
for some $\epsilon_{n}$ such that $\limsup_{n\rightarrow\infty}\epsilon_{n}=0$.

Set $U=M_{1},V=M_{2}$ and follow similar steps to Subsection \ref{subsec:Converse},
then we have

\begin{equation}
C_{\beta}^{(D)}(X;Y)=\lim_{n\rightarrow\infty}\inf_{P_{U|X^{n}}P_{V|Y^{n}}:\rho_{m}(X^{n};Y^{n}|UV)\leq\beta}\frac{1}{n}{\displaystyle I(X^{n}Y^{n};UV)}\leq R.
\end{equation}
Hence
\begin{equation}
R_{DCIE}(\beta)\geq C_{\beta}^{(D)}(X;Y).
\end{equation}
Combining this with the achievability of $C_{\beta}^{(D)}(X;Y)$ gives
us
\begin{equation}
R_{DCIE}(\beta)=C_{\beta}^{(D)}(X;Y).
\end{equation}

Now we remain to show
\begin{equation}
C_{\beta}^{(D)}(X;Y)\geq C_{\beta}^{(D,LB)}(X;Y).
\end{equation}

Consider $P_{U|X^{n}}P_{V|Y^{n}}$ such that $\rho_{m}(X^{n};Y^{n}|UV)\leq\beta$
and $\frac{1}{n}{\displaystyle I(X^{n}Y^{n};UV)}\leq R$. Then the
following equations hold.
\begin{align}
 & U\rightarrow X_{T}\rightarrow Y_{T},\\
 & X_{T}\rightarrow Y_{T}\rightarrow V,\\
 & \rho_{m}(X_{T},Y_{T}|UVT)\leq\rho_{m}(X^{n};Y^{n}|UV),\\
 & \rho_{m}(UX_{T};VY_{T}|T)\leq\rho_{m}(UX^{n};VY^{n})=\rho_{m}(X^{n};Y^{n})=\rho_{m}(X;Y),
\end{align}
and
\begin{align}
I_{Q}(X^{n}Y^{n};UV) & ={\displaystyle \sum_{i=1}^{n}I_{Q}(X_{i}Y_{i};UV|X^{i-1}Y^{i-1})}\\
 & ={\displaystyle \sum_{i=1}^{n}I_{Q}(X_{i}Y_{i};UVX^{i-1}Y^{i-1})}\\
 & =nI_{Q}(X_{T}Y_{T};UVX^{T-1}Y^{T-1}|T)\\
 & =nI_{Q}(X_{T}Y_{T};UVX^{T-1}Y^{T-1}T)\\
 & \geq nI_{Q}(X_{T}Y_{T};UV|T)\\
 & =nI_{Q}(XY;UV|T),
\end{align}
where $T$ is a time-sharing random variable uniformly distributed
$[1:n]$ and independent of all other random variables, and $X:=X_{T},Y:=Y_{T}$.
Therefore,
\begin{equation}
C_{\beta}^{(D)}(X;Y)\geq C_{\beta}^{(D,LB)}(X;Y).
\end{equation}
This completes the proof.

\end{document}